\documentclass[twocolumn,
english,
aps,
pra,
10pt,
longbibliography,
amsmath,
amssymb,
floatfix
]{revtex4-2}

\usepackage{graphicx}
\usepackage{booktabs}
\usepackage{multirow}
\usepackage{natbib}
\usepackage{array}
\usepackage{ragged2e}
\usepackage{microtype}
\usepackage{natbib}
\usepackage{xcolor}
\usepackage{physics}
\usepackage{nicematrix}
\usepackage{lmodern}

\usepackage[colorlinks=true,urlcolor=blue,citecolor=blue,linkcolor=blue]{hyperref}
\usepackage{capt-of}

\newcolumntype{L}[1]{>{\RaggedRight\arraybackslash}p{#1}}
\newcolumntype{C}[1]{>{\Centering\arraybackslash}p{#1}}

\newcommand{\NR}{\textit{NR}}
\newcommand{\NA}{\textit{N/A}}
\newcommand{\derived}{\textsuperscript{\textdagger}}

\begin{document}

\makeatletter
\renewcommand*{\@fnsymbol}[1]{\ensuremath{\dagger}}
\makeatother

\title{Learning angles for quantum approximate optimization}

\author{Johannes Jakubik}
\thanks{Corresponding authors: \url{johannes.jakubik1@ibm.com}, \url{deg@zurich.ibm.com}.}
\affiliation{IBM Research}

\author{Isabelle Wittmann}
\affiliation{IBM Research}

\author{Zaheed Gaffoor}
\affiliation{IBM Research}

\author{Gciniwe Simphiwe Baloyi}
\affiliation{IBM Research}

\author{Craig Mahlasi}
\affiliation{IBM Research}

\author{Etienne Vos}
\affiliation{IBM Research}

\author{Thomas Brunschwiler}
\affiliation{IBM Research}


\author{Daniel J. Egger}
\thanks{Corresponding authors: \url{johannes.jakubik1@ibm.com}, \url{deg@zurich.ibm.com}.}
\affiliation{IBM Research}

\begin{abstract}
The Quantum Approximate Optimization Algorithm (QAOA) and its variants are leading algorithms to help solve challenging combinatorial optimization problems.
QAOA maps these problems to Ising Hamiltonians such that their ground state is the desired solution.
Candidate solutions are then drawn from a variational quantum circuit.
The performance of QAOA hinges on a set of variational angles. 
However, finding good angles is an instance-specific and computationally expensive non-convex classical optimization task.
Here, we learn the mapping from the problem instance to high-quality QAOA angles.
Our learned models replace the per-instance optimization, that we call the solver, with a single forward pass that is several orders of magnitude faster than the solver with a near identical quality.
Our models further generalize to unseen topologies such as Barab\'asi--Albert and 2D grid graphs.
Experiments on quantum hardware confirm that QAOA samples with angles produced by our open-source models are indistinguishable from solver angles.
Crucially, models sensitive to the topology of the problem instance increase the quality of the QAOA angles.
Our work amortizes a major bottleneck in QAOA by an up-front data generation and training cost.
Importantly, this cost can be amortized over time in settings where many similar problems are optimized on a regular basis, as is often the case in applications for industry.
\end{abstract}

\maketitle

\section{Introduction}

Combinatorial optimization underlies problems throughout logistics, finance, and the physical sciences~\cite{Dantzig1959, Markowitz1952, Mohseni2022}. 
Yet, many of the most important instances are NP-hard and remain intractable for exact classical methods at scale~\cite{Karp1972, Garey1979}.
By processing information with the laws of quantum mechanics, quantum computers offer new algorithmic tools to tackle such problems~\cite{Abbas2024}.
The quantum approximate optimization algorithm (QAOA)~\cite{Farhi2014} aims to solve combinatorial problems by finding the ground state of a corresponding Hamiltonian $H_C$~\cite{Lucas2014}.
QAOA samples candidate solutions from a quantum state built by alternating $p$ layers of $e^{-i\gamma_k H_C}$ and $e^{-i\beta_k H_M}$. 
Here, $H_M$ is the mixer Hamiltonian, whose easy-to-prepare ground state is the initial state.
Crucially, the performance of QAOA depends on the $2p$ angles $\boldsymbol{\beta}$, $\boldsymbol{\gamma}$.
Finding good angles is, however, expensive: the optimization landscape is non-convex and rugged, and locating high-quality angles can be NP-hard~\cite{bittel2021training}.

This bottleneck motivates many QAOA angle selection methodologies~\cite{Guo2026}, ranging from analytic fixed-angle conjectures~\cite{Wurtz2021} to a wide variety of machine-learning (ML) strategies to predict, transfer, extrapolate, or optimize $(\boldsymbol{\beta}, \boldsymbol{\gamma})$.
Methods that directly predict instance-specific QAOA angles are often confined to small, classically simulable graphs of up to roughly $20$ nodes and shallow depths. 
Other approaches reach larger graph sizes or depths through angle transfer, depth extrapolation, or continued instance-specific optimization. 
Existing ML approaches therefore address distinct aspects of the broader problem of learning QAOA angles under substantially varying experimental settings, see App.~\ref{app:rw_settings}. 
Therefore, it remains an open question whether learned models can directly predict high-quality angles across a broad range of problem sizes and topologies without further per-instance optimization.

To address this question, we frame QAOA angle selection as a supervised, graph-level regression problem. 
For a given target depth $p$, a single neural network maps a weighted graph representing the optimization problem directly to the $2p$ QAOA angles.
This replaces costly per-instance angle optimization with a single inexpensive forward pass. 
The following points summarize our contributions:
(i)~We train and evaluate a wide range of ML models on a dataset of QAOA angles comprising 12\,191 weighted \textsc{MaxCut} instances spanning random regular, Erd\H{o}s--R\'enyi, and heavy-hex topologies with up to $144$ nodes. 
By rescaling $\boldsymbol{\gamma}$ on a per-instance basis, we remove the weight-dependent scale of $H_C$ and enable a single model to learn across graph families and weight distributions.
(ii)~Across circuit depths $p \in \{1, \dots, 5\}$, the predicted angles achieve approximation ratios close to those obtained by COBYLA~\cite{Powell1994} without post-prediction optimization.
This reduces the time to obtain $\boldsymbol{\beta}$ and $\boldsymbol{\gamma}$ by several orders of magnitude, yielding a favorable trade-off between solution quality and runtime. 
(iii)~A single trained model generalizes to graphs substantially larger than those seen in training (up to $200$ nodes) and to unseen topologies such as Barab\'asi--Albert and 2D-grid graphs, without any per-instance retraining or fine-tuning. 
(iv)~We evaluate our predicted angles on IBM superconducting quantum hardware for heavy-hex graphs up to depth $p=3$ and open-source our training and inference code together with pretrained model weights.

The rest of the paper is organized as follows. Sec.~\ref{sec:related-work}  reviews ML approaches for QAOA and provides a high-level overview of our method. 
Sec.~\ref{sec:qaoa_fixed} introduces QAOA and the analytic fixed-angle baseline. 
Sec.~\ref{sec:dl_qaoa} describes the data generation process and ML models. 
Sec.~\ref{sec:results} presents and discusses the results.
We conclude in Sec.~\ref{sec:conclusion}.

\begin{figure*}[ht]
    \centering
    \includegraphics[width=\linewidth]{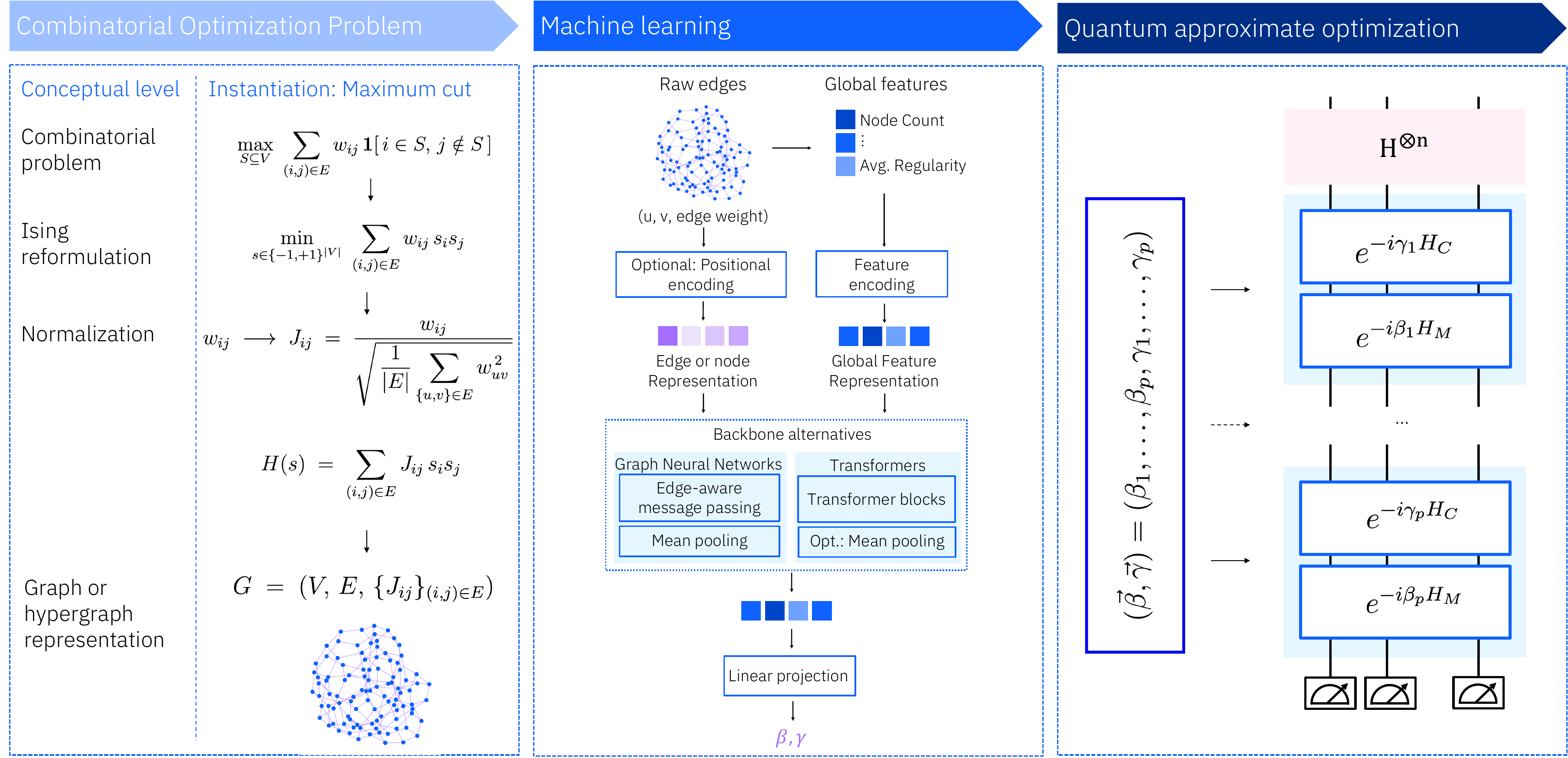}
    \caption{Learning angles for QAOA: In Phase~(1), we cast the optimization problem into a (hyper-)graph representation via a normalized Hamiltonian, which allows us to train a deep learning model in Phase~(2) across varying graph types. Deep learning models mostly rely on information of the raw graph topologies as well as aggregated global features. Backbones can be distinguished into Graph Neural Network-type models and Transformer-type models. In Phase~(3), we describe the inference, where we derive angles $(\beta, \gamma)$ via single forward passes and test their quality through simulation and on quantum hardware.}
    \label{fig:overview}
\end{figure*}

\section{Related Work}
\label{sec:related-work}

\citet{Guo2026} group QAOA angle-setting strategies into physics-inspired methods, angle transfer, iterative methods, and ML methods. 
We focus on ML approaches, in which a learned component is trained on previously solved instances or on the target instance itself to guide angle setting.
As these approaches can overlap with the other categories, we organize them by the role of the learned component: direct angle prediction, angle selection and transfer, depth extrapolation, and learned guidance of iterative optimization. 

\emph{Direct Angle Prediction} methods use ML to infer QAOA angles directly from a problem instance. 
Most approaches use supervised regressors that map a graph to an angle vector. 
Early work applies Multilayer Perceptrons (MLPs) to flattened adjacency matrices~\cite{amosy2024iteration}, which are sensitive to node relabeling. 
Later approaches use Graph Neural Networks (GNNs) that can operate on the graph itself~\cite{deshpande2022capturing, liang2024graph}. 
QSeer~\cite{jiang2025qseer} further incorporates physics-inspired constraints and supports weighted graphs.
Generative approaches instead learn conditional distributions over QAOA angles or layer structures. 
Meng \emph{et al.}~\cite{meng2025conditional} use a denoising diffusion model to generate initial angles at a fixed depth, whereas QAOA-GPT~\cite{Tyagin2025} uses an autoregressive transformer to jointly generate QAOA layers and their associated angles.
Prior direct predictors have largely been trained and evaluated on classically simulable graphs with at most 20 nodes and up to $p=4$.

\emph{Angle Transfer} methods exploit the concentration of optimal angles across similar graphs by reusing angles from suitable donor instances~\cite{galda2021transferability, galda2023similarity}. 
While classical approaches use fixed angles or handcrafted similarity measures, ML-based methods identify donors through learned graph embeddings and clustering~\cite{moussa2022unsupervised, xu2025qaoa}. Such methods can approach full optimization at lower cost and have been evaluated on graphs with up to 100 nodes and depths up to $p=10$~\cite{falla2024graph}, but rely on angles from related instances.

\emph{Depth-Based Extrapolation} methods use ML to infer angles for deeper QAOA circuits from solutions at lower depths. 
Building on iterative heuristics such as INTERP and FOURIER~\cite{Zhou2020}, learning-based approaches include one-shot regressors that map $p=1$ angles to higher depths~\cite{alam2020accelerating} and autoregressive models that predict angles layer by layer~\cite{xie2023quantum}. 
Across these methods, the inferred higher-depth angles are used to initialize subsequent classical optimization at the target depth.

\emph{Learned Optimization} methods use ML to guide an iterative search rather than directly outputting QAOA angles. 
Reinforcement-learning approaches learn angle-update policies or edge-selection strategies for recursive QAOA variants~\cite{khairy2020learning, yao2020policy, wauters2020reinforcement, patel2024reinforcement}.
Recurrent models learn optimization trajectories from previously solved instances~\cite{chandarana2023meta, chen2025learning}, while surrogate-based methods iteratively select angle configurations using predicted objective values~\cite{cheng2024quantum}. 
Neural QAOA$^2$~\cite{Zheng2026} predicts graph partitions and topology-aware initial angles for divide-and-conquer QAOA, using a differentiable surrogate.
Other approaches use Monte Carlo Tree Search to explore the parameter landscape~\cite{agirre2025monte}.
Related configuration-selection methods predict QAOA performance rather than angle updates. 
QAOA-Predictor~\cite{Coelho2026} estimates the success probability of linear-ramp QAOA from the problem instance, ramp parameter, and circuit depth, enabling selection of a suitable parameterization and minimal depth.
Most such methods have been evaluated on problems with up to 30 nodes. Neural QAOA$^2$ considers problems with up to 1,000 variables by decomposing them into subproblems of at most 10 qubits~\cite{Zheng2026}.

Across these categories, evaluation settings differ, making direct comparison of approaches challenging. Most results rely on classical simulation, with notable hardware exceptions including DARBO~\cite{cheng2024quantum} and utility-scale benchmarks of transferred angles~\cite{Guo2026}.  Appendix~\ref{app:rw_settings} provides a comparison of evaluation settings across related work.

Our work falls under direct angle prediction. We infer angle vectors for fixed-depth standard QAOA directly from the problem graph, using a single forward pass and no instance-specific refinement. 
We compare various model architectures under a common protocol, train them across several graph families, circuit depths up to $p=5$ and evaluate graphs with up to 200 nodes. Two additional graph families test out-of-distribution generalization. Crucially, we benchmark the inferred angles against known fixed-angle baselines with strong performance guarantees~\cite{Wurtz2021} and validate the angles for heavy-hex instances on quantum hardware.

\section{QAOA and fixed angles}
\label{sec:qaoa_fixed}

QAOA approximates the ground state of a cost Hamiltonian \(H_C\) that encodes a classical optimization problem. 
Depth $p$ QAOA samples from the variational ansatz state
\begin{align}
\lvert \psi_p(\boldsymbol{\beta}, \boldsymbol{\gamma}) \rangle
    = \prod_{k=1}^{p}
      \left(
        e^{-i \beta_k H_M}
        e^{-i \gamma_k H_C}
      \right)
      \lvert + \rangle^{\otimes n}.
\end{align}
Here, \(H_M = -\sum_{j=1}^n X_j\) is the mixing Hamiltonian and \(\boldsymbol{\beta} = (\beta_1,\dots,\beta_p)\), \(\boldsymbol{\gamma} = (\gamma_1,\dots,\gamma_p)\) are real variational parameters.
These angles are optimized to minimize the expected energy
\begin{align}\label{eq:qaoa_energy}
E(\boldsymbol{\beta}, \boldsymbol{\gamma})
    = \langle \psi_p(\boldsymbol{\beta}, \boldsymbol{\gamma}) \rvert
      H_C
      \lvert \psi_p(\boldsymbol{\beta}, \boldsymbol{\gamma}) \rangle,
\end{align}
to drive the ansatz state toward the ground state of \(H_C\).

As a benchmark, we consider the weighted \textsc{MaxCut} problem. 
Here, we seek a partition of the vertex set \(V\) of a graph \(G = (V,E)\) such that the total weight of edges crossing the cut is maximized. 
For binary decision variables \(x_i \in \{0,1\}\) indicating the side of the cut containing vertex \(i\), the objective is
\begin{align}
\max_{x \in \{0,1\}^n}
     \sum_{(i,j) \in E}
      w_{ij}\, \left[x_i(1-x_j)+x_j(1-x_i)\right],
\end{align}
where \(w_{ij}\) is the weight of edge $(i,j)$.
To solve weighted \textsc{MaxCut} on a quantum computer, we map the binary 
variables \(x_i\) to spin variables $z_i=1-2x_i$,
and promote the spin variables $z_i$ to Pauli-$Z$ operators $Z_i$. 
Using the ground-state convention above, this yields the weighted \textsc{MaxCut} Hamiltonian
\begin{align}
H_C
    = \frac{1}{2}
      \sum_{(i,j) \in E}
      w_{ij}\, (Z_i Z_j-1).
\end{align}

We measure the quality of angles $(\boldsymbol{\beta},\boldsymbol{\gamma})$ by the \emph{approximation ratio}
\begin{align}
\mathrm{AR}
    = \frac{
        \langle C(x)\rangle - C_{\min}
      }{
        C_{\max} - C_{\min}
      },
\end{align}
where \(C_{\max}\) and \(C_{\min}\) are the maximum and minimum cut values, respectively, obtained with CPLEX, a classical optimizer. 
Furthermore, $\langle C(x)\rangle$ is the average cut value of the candidate solutions $x$ sampled from $\ket{\psi_p(\boldsymbol{\beta},\boldsymbol{\gamma})}$.
The approximation ratio therefore lies in the interval \([0,1]\). 

For a given graph, generated using NetworkX~\cite{hagberg2007exploring}, the angles \(\boldsymbol{\beta}\) and \(\boldsymbol{\gamma}\) can be obtained through a variety of strategies~\cite{Guo2026}. 
We find instance specific angles by first performing a two-dimensional $40\times40$ grid scan in the interval $[0,\pi/2]^2$ over $(\beta_1, \gamma_1)$ at $p=1$.
This is crucial because it avoids multi-modal QAOA angle distributions, e.g., caused by reflections and $\pi$-shifts, that would complicate model training.
Next, we refine the resulting angles with COBYLA set to a small initial trust region of $\texttt{rhobeg}=0.05$ and a maximum of 100 iterations.
At depth $p=1$ we evaluate the energy  exactly~\cite{Egger2021}.
Finally, we iterate the linear interpolation of Zhou \emph{et al.}~\cite{Zhou2020} from $p=1$ to $5$.
We compute the energy at $p>1$ with Pauli Propagation with maximum Pauli weights and coefficients of 4 and $10^{-5}$, respectively~\cite{Rall2019, Rudolph2025}. 
This avoids a full statevector simulation at the cost of an approximate energy evaluation.
We refer to this approach as the \emph{solver}.
The \emph{solver} results in high AR but requires a long runtime.

\begin{figure}
    \centering
    \includegraphics[width=0.95\columnwidth, clip, trim=0 340 325 0]{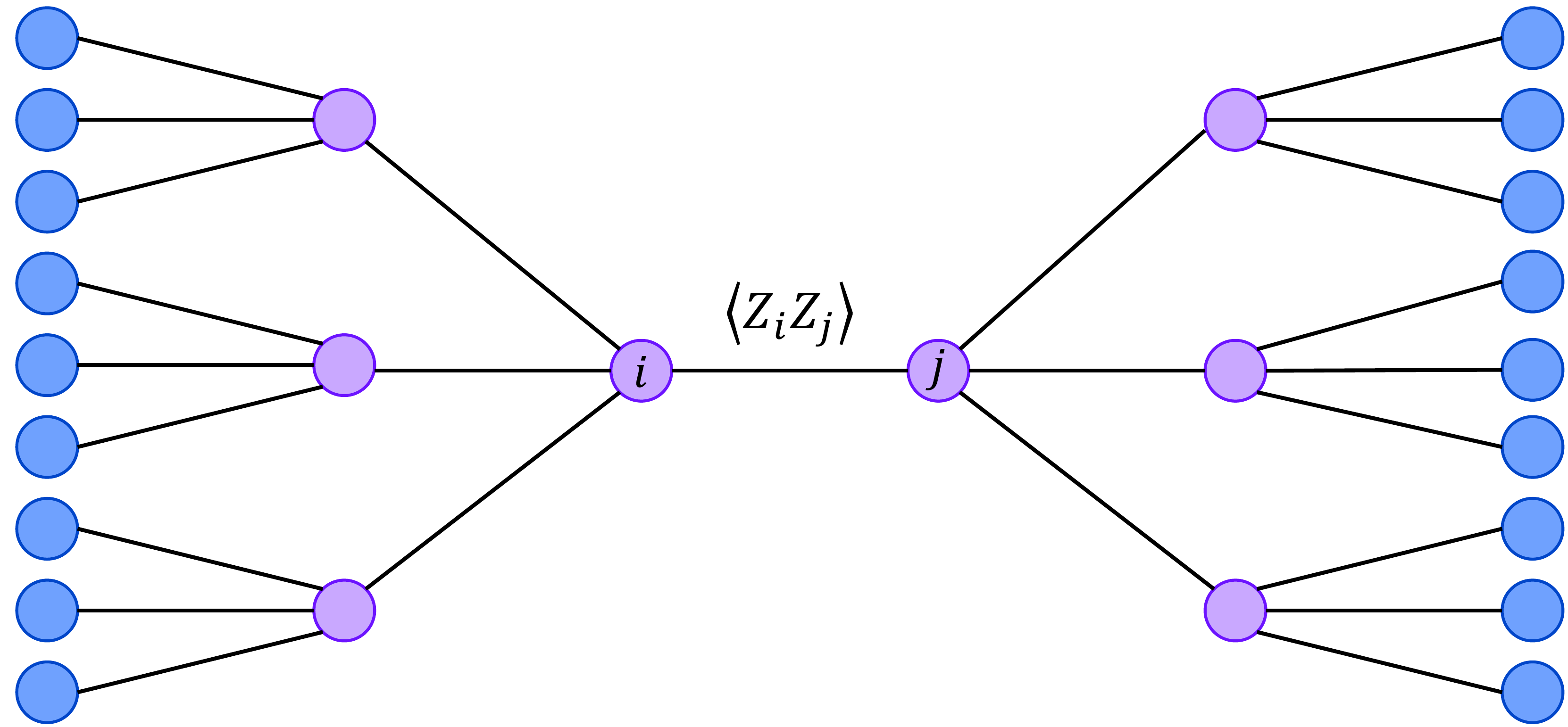}
    \caption{Tree graph underlying the fixed-angles design for random four-regular graphs.
    The purple nodes show the tree seen by depth-one QAOA.
    The purple and blue nodes show the tree seen by depth-two QAOA.}
    \label{fig:fixed_angles_tree}
\end{figure}

By contrast to the \emph{solver}, a \emph{fixed‑angle} approach derives high‑quality QAOA angles by exploiting the local structure seen by QAOA on random regular graphs~\cite{Wurtz2021}. 
For a random $d$-regular graph with no loops smaller than $2p+1$, the local subgraphs relevant at QAOA depth $p$ are identical trees. 
The energy in Eq.~(\ref{eq:qaoa_energy}) is thus written as  $|E|F_{d,\mathrm{tree}}(\boldsymbol{\beta},\boldsymbol{\gamma})$, where $F_{d,\mathrm{tree}}$ denotes the energy associated with the corresponding tree graph, as illustrated in Fig.~\ref{fig:fixed_angles_tree} for $p=1, 2$. 
QAOA angles optimal to tree graphs with non-leaf nodes of degree $d$ lower bound the approximation ratio of random $d$-regular graphs.
Since the quantum circuits underlying $\langle Z_iZ_j\rangle$ on tree graphs can be evaluated with tree tensor networks, fixed angles have been computed, and stored in a database, for depths up to $p=11$ for random three-regular graphs.
With such a database, obtaining the fixed angles reduces to a constant-time lookup.
For graphs $G$ that are not regular we use the fixed angles of the $d$-regular graph in the database with the closest degree to the average node degree of $G$. For heavy-hex graphs, we use fixed-angles that have been specifically tailored to this graph topology for up to $p=3$.

\section{Deep learning for QAOA}
\label{sec:dl_qaoa}

The fixed-angle approach relies on the local subgraph structures that characterize the QAOA circuit at a given depth. For random regular graphs, these structures are repetitive, identical tree structures, allowing a single set of tree-derived angles to generalize across graph instances. General graphs instead contain a heterogeneous distribution of local subgraph structures and edge weights, whose individual contributions and relative frequencies collectively determine the QAOA energy. This suggests that a mapping from graph structure to high-quality QAOA angles can be learned from a diverse set of solved instances, avoiding the need for per-instance optimization.

We therefore formulate QAOA angle prediction as supervised graph-level regression. Given a weighted graph $G=(V,E)$, we learn a mapping to the $2p$ angles
\begin{align}
f_\theta(G) = \left(\beta_1,\ldots,\beta_p,\gamma_1,\ldots,\gamma_p\right).
\end{align}
For each circuit depth $p$, we train a single model across multiple graph families, with the aim of generalizing across graph sizes, edge weight distributions, and both seen and unseen topologies. 
The remainder of this section describes the generation of optimized QAOA training labels and the graph-based models used to learn this mapping.

\subsection{Data Generation}
\label{sec:data_generation}

\begin{figure*}[ht]
    \centering
    \includegraphics[width=1\linewidth]{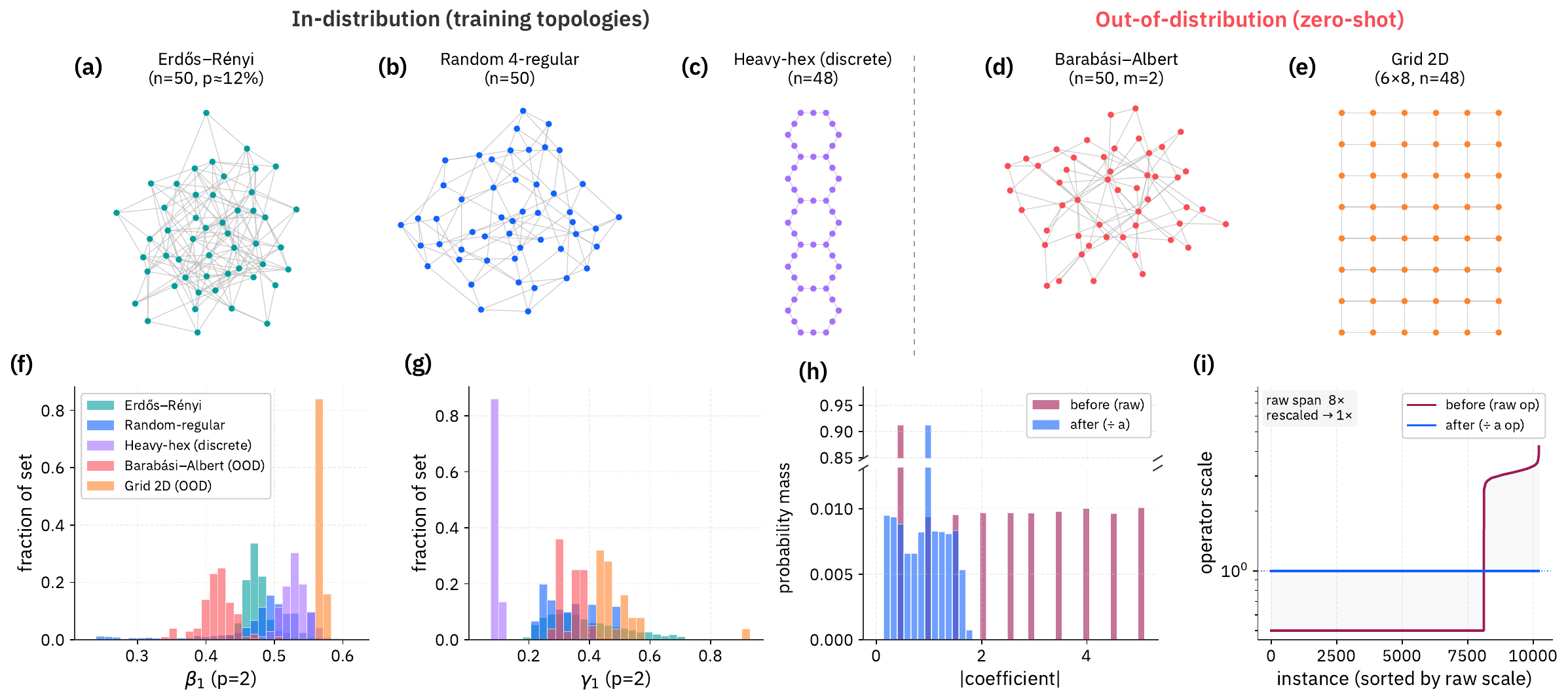}
    \caption{
    Panels (a), (b), and (c) exemplify in-distribution Erd\H{o}s--R\'enyi, random-regular, and heavy-hex graphs, respectively.
    Panels (d) and (e) exemplify out-of-distribution Barabási–Albert and 2D grid graphs, respectively.
    Panels (f) and (g) show the distribution of the $\beta_1$ and $\gamma_1$ angles, respectively, obtained from the \emph{solver} at $p=2$ prior to rescaling $\boldsymbol{\gamma}$.
    (h) Distribution of the $\boldsymbol{\gamma}$ angles before and after rescaling with Eq.~(\ref{eq:rescaling}).
    (i) Per-instance cost-operator scale factor $a$ across all five graph topologies.}
    \label{fig:graph_topologies}
\end{figure*}

Supervised training requires a sufficiently large dataset of graph instances paired with high-quality QAOA angles as labels. 
We generate training examples across four graph families, namely, unweighted random $d$‑regular graphs, Erd\H{o}s--R\'enyi graphs, and Heavy-Hex graphs with edges weights uniformly drawn from either $[-1, 1]$ or $\{1, 2, ..., 10\}$, see Fig.~\ref{fig:graph_topologies}(a)-(c).
Table~\ref{tab:graph_instances_dist} summarizes the graph properties and number of instances for each setting.

\begin{table}
    \centering
\caption{Distribution of the graph instances for the four graph families. 
Here, $N_I$ is the number of graph instances in the dataset.
The heavy-hex graphs have $m\in \{1,2, ..., 7\}$ rows and $n\in \{1,2, 3\}$ columns. 
The number of rows and columns of the 2D grid graphs are chosen from $\{2,...,10\}$.
}
\label{tab:graph_instances_dist}
    \begin{tabular*}{\columnwidth}{@{\extracolsep{\fill}}l c c r@{}}\hline\hline
    Family & $N_I$&  Size &  Properties \\\hline
    \multicolumn{4}{l}{In distribution}\\\hline
    random $d$-regular & 1,561&  10 - 100&  $d\in\{3, \dots, 10\}$\\
    {Erd\H{o}s--R\'enyi} &6,430&  10 - 100 & $p_i\in\{0.05, \dots, 0.12\}$\\
    Heavy-hex (uniform) & 2,100& 10 - 144 & $w_{ij}\in[-1, 1]$\\ 
    Heavy-hex (discrete) & 2,100& 10 - 144& $w_{ij}\in\{1, 2, ..., 10\}$\\\hline 
    \multicolumn{4}{l}{Out of distribution}\\\hline
    Barabási–Albert & 100 & 10 - 50 & $m\in\{2,3\}$ \\
    2D grid & 25 & 4 - 100 & $m, n \in \{2 , \dots, 10\}$\\
    {Erd\H{o}s--R\'enyi} & 20 & 150, 200 & $p_i = 0.05$ \\
    \hline\hline
    \end{tabular*}
\end{table}

\begin{figure}
    \centering
    \includegraphics[width=\columnwidth]{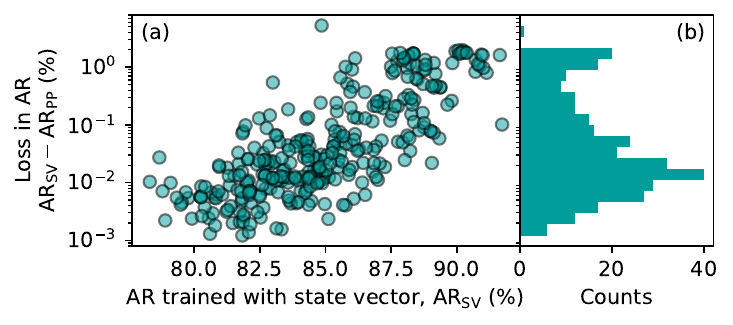}
    \caption{Quality validation of the QAOA angle training with Pauli propagation at $p=2$ and 20 nodes.
    (a) AR loss of the angles optimized with Pauli propagation versus the AR of angles optimized with a state vector.
    Each marker is one of either 160 random $d$-regular graphs with $d\in\{3, 4, ..., 10\}$ or 160 Erd\H{o}s--R\'enyi graphs with $p_i\in\{5\%, 6\%, ..., 12\%\}$.
    The ARs of all the trained angles are evaluate with a state vector.
    (b) Distribution of the loss in AR.
    }
    \label{fig:validation}
\end{figure}

We obtain the corresponding QAOA training labels $\beta_1$, ..., $\beta_p$ and $\gamma_1,...,\gamma_p$ up to $p=5$ with the \emph{solver} described in Sec.~\ref{sec:qaoa_fixed}. 
Since Pauli propagation provides an approximate energy evaluation, we validate its accuracy on instances with 20 nodes. 
Here, the approximation ratios obtained using QAOA angles optimized with Pauli propagation are nearly identical to those obtained from angles optimized on the same graph instances using an exact state‑vector energy evaluator, see Fig.~\ref{fig:validation}. 
Additional implementation details regarding the data generation, distributions of training examples, and the data splits are provided in App.~\ref{app:dataset_details}.

In addition to the in-distribution graphs we also generate out of distribution graphs to evaluate the extrapolation performance of the models.
Erd\H{o}s--R\'enyi graphs with 150 and 200 nodes allow us to test model extrapolation to unseen graph sizes.
Barabási–Albert (BA) and 2D grid graphs allow us to test models on unseen graph topologies.
In particular, when building the BA graphs we attach each new node to $m\in\{2,3\}$ existing nodes to ensure that the average node degree is in the fixed angles database. 
Moreover, the edge weights for both BA and 2D grid graph families were sampled uniformly at random from the discrete set $\{-2, -1, 1, 2\}$.

\subsection{Target normalization}

The QAOA applies the unitary $e^{-i\gamma H_C}$, such that a global rescaling of the cost Hamiltonian results in an inverse rescaling of the corresponding $\gamma$ angles. 
Graph instances with large edge weights therefore generally require small $\gamma$ values. 
The scale of $H_C$ can vary substantially since it depends on the problem at hand. 
This introduces an additional variation in the regression targets. 
We reduce this dependency through a per-instance rescaling of the cost operator, i.e., $H_C\to a^{-1}H_C$, where
\begin{equation}\label{eq:rescaling}
a = \sqrt{\sum_{\mathrm{order}} \overline{w^2}_{\mathrm{order}}},
\end{equation}
following Eq.~(8) of Ref.~\cite{Sureshbabu2024}.
Here, $\overline{w^2}_{\mathrm{order}}$ denotes the mean squared magnitude of the cost operator's Pauli coefficients grouped by locality order. 
For example, \textsc{MaxCut} only has second order terms such that $\smash{a=\sqrt{\overline{w^2}_2}}$ with $\smash{\overline{w^2}_2=|E|^{-1}\sum_{(i,j)\in E}w_{ij}^2}$ on a weighted graph. 
We use $\widetilde{\boldsymbol{\gamma}}=\boldsymbol{\gamma} a$ as the training target, while the $\boldsymbol{\beta}$ angles remain unchanged. 
At inference time, the predicted $\widetilde{\boldsymbol{\gamma}}$ values are divided by $a$ to recover the angles used for circuit evaluation.
On the training data, the rescaled $\widetilde{\boldsymbol{\gamma}}$ have a smaller range than  $\boldsymbol{\gamma}$, see Fig.~\ref{fig:graph_topologies}(g)-(i).
We find that this rescaling is crucial to train a single model on the different graph families.

\subsection{Machine learning models}
\label{sec:ml_models}

We evaluate seven machine learning architectures for QAOA angle prediction. 
As a non-structural baseline, we use a multi-layer perceptron (MLP) that takes global graph features as input, including the number of nodes and edges, mean degree, and degree standard deviation. 
A sequence of fully connected layers maps these features directly to the $2p$ output angles, providing a reference for how well QAOA angles can be predicted from coarse graph properties alone. 
All global features are standardized using statistics from the training set.

\begin{figure}[h!]
    \centering
    \includegraphics[width=1.0\columnwidth]{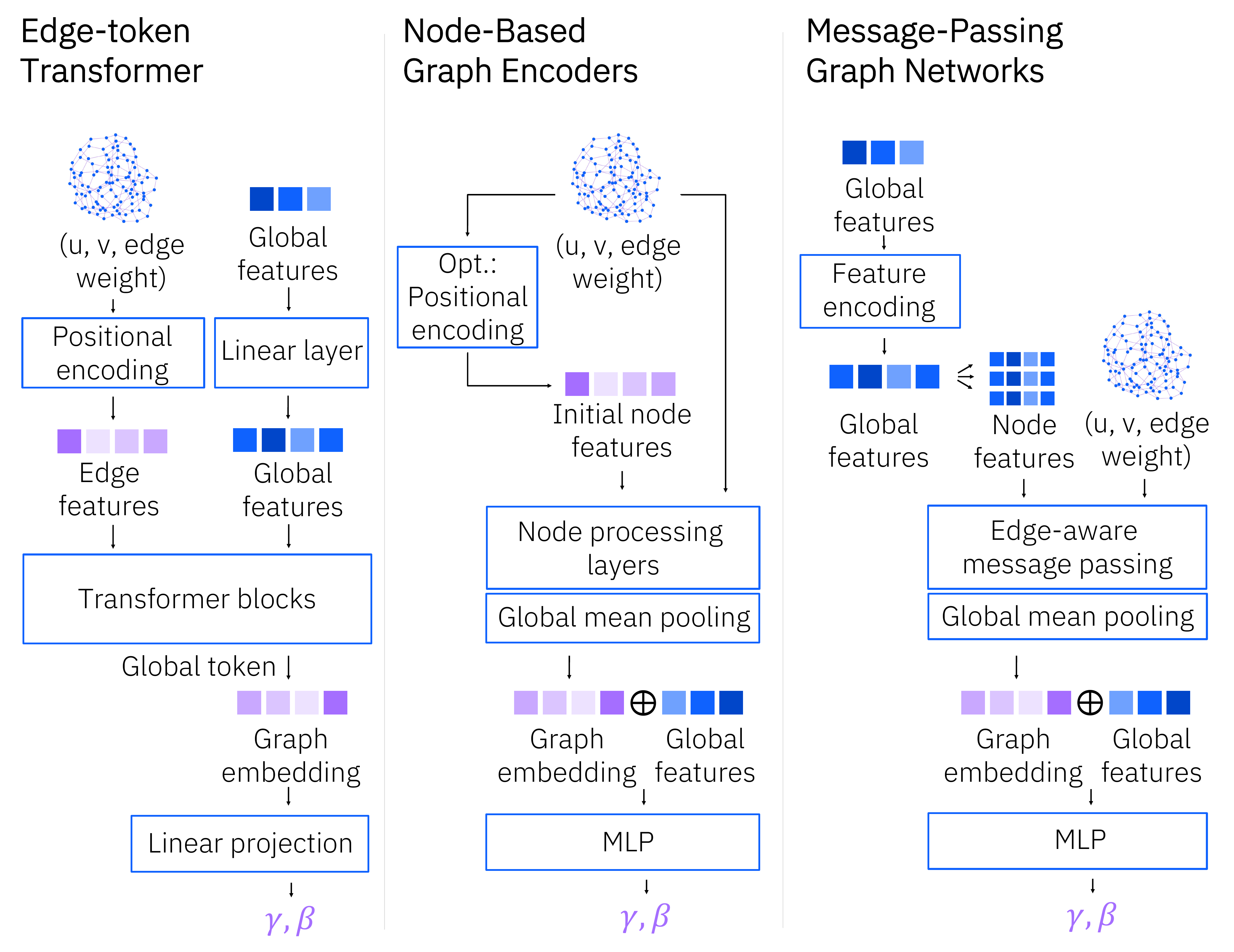}
    \caption{Schematic overview of the three classes of graph-based models used for QAOA angle prediction. The models differ in how graph structure and global graph features are represented, processed, and combined before predicting the QAOA angles.}
    \label{fig:model_overview_main}
\end{figure}

The remaining models additionally use the graph structure and edge weights. 
For comparison, we group our implementations into three classes according to how this information is represented and processed, see Fig.~\ref{fig:model_overview_main}.
Edge-token transformers represent each edge through its endpoints and weight as a separate token, together with a global token constructed from the global graph features. 
The tokens are processed jointly using self-attention~\cite{vaswani2017attention}, after which the global token is projected onto the output angle vector.
Node-based graph encoders instead represent the graph through node features, such as node degree, and process them together with the graph connectivity using either Graph Transformer~\cite{dwivedi2021generalizationtransformernetworksgraphs} or graph convolutional~\cite{morris2019weisfeiler} layers. The resulting node representations are averaged, concatenated with the global graph features, and mapped to the output angles using an MLP.
Message-passing graph networks similarly operate on node representations and graph connectivity, but update node features by passing information between neighboring nodes along the edges~\cite{gilmer2017neural}. In our models, the node representations are initialized from the global graph features and updated through edge-aware message-passing layers before being pooled and mapped to the output angles.

Within these three classes, we evaluate an Edge Transformer (EdgeTF) and a diffusion-based transformer (DiffTF), a Graph Transformer (GraphTF) and graph convolutional network (GCN), and a Graph Neural Network (GNN) and Graph Isomorphism Network (GIN), respectively. Further architectural and implementation details are provided in App.~\ref{app:model_details}. This comparison tests the value of coarse graph statistics for QAOA angle prediction and how performance depends on the representation of graph structure.

\subsection{Model training}

For each architecture and circuit depth, training is performed jointly across all four graph families described in Sec.~\ref{sec:data_generation}. 
We use fixed training, validation, and test splits that are shared across all model architectures, see App.~\ref{app:dataset_details}. 
The mean-squared error (MSE) between the predicted and target angle vectors $(\boldsymbol{\beta},\tilde{\boldsymbol{\gamma}})$ is the training objective.
To identify suitable hyperparameters across models, we perform a grid search over combinations of learning rates \{1e-4, 5e-4, 1e-3, 1e-2\} and batch sizes \{8, 32\} based on validation performance. 
We use the Adam optimizer~\cite{kingma2017adam} along with a selected learning rate of $1\times10^{-3}$ and batch size of 64, a 5-epoch linear warm-up, and cosine annealing for the remaining epochs. 
Throughout training the gradient norms are clipped to 1.0. 
The models are trained for a maximum of 100 epochs with early stopping based on validation loss. 
To obtain robust performance estimates, we train each model using five independent random seeds (42--46). 
Unless stated otherwise, all reported results are averaged over these five training runs.

\section{Results}
\label{sec:results}

\begin{figure*}[t]
\centering
\includegraphics[width=\linewidth]{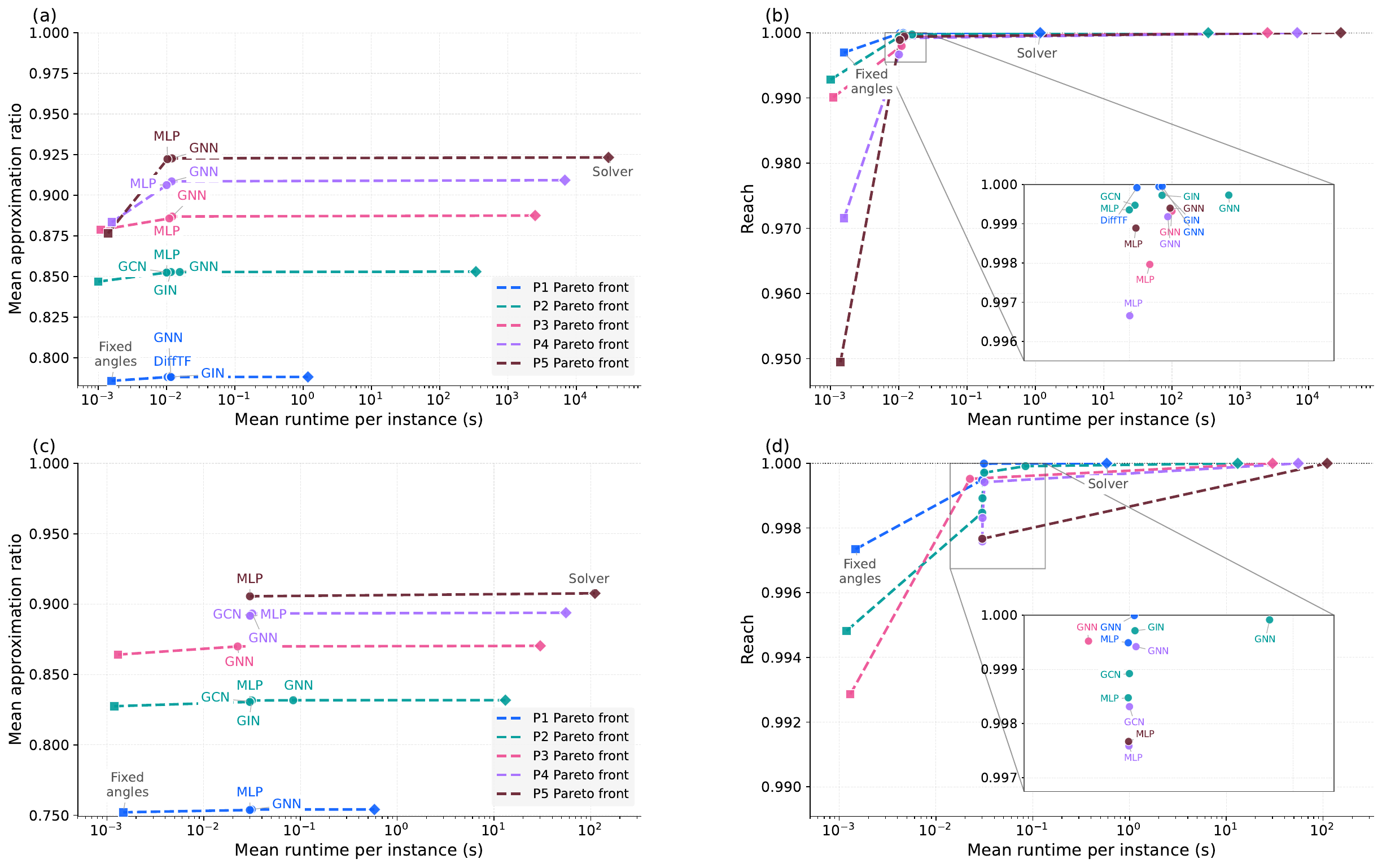}
\caption{
    Pareto front of the trade-off between method runtime and AR for $p=1$ to $5$ averaged over the Test Set. 
    Panels (a) and (b) show the AR and Reach, respectively, for Erd\H{o}s--R\'enyi and random-regular graphs.
    Panels (c) and (d) show the AR and Reach, respectively, for the heavy-hex graphs with uniform and discrete weighted.
    We only show methods that are on the Pareto front.
}
\label{fig:pareto}
\end{figure*}

First, we characterize runtime--accuracy trade-offs by locating our learned models relative to the fixed-angles baseline and the upper bound formed by the \emph{solver}. 
Second, we assess predictive quality across circuit depths. 
Third, we probe generalization to graph sizes and topologies not seen during training. 
Fourth, we aim to better comprehend limitations of MSE-style proxy training objective compared to the underlying energy landscape. 
Finally, we investigate the quality of predicted angles on real quantum hardware. 

\subsection{Pareto fronts}
\label{sec:res_pareto}

The central advantage of learned angle inference is the cost of obtaining angles. 
We study the Pareto frontier of approximation ratio against runtime for 
(i)~the \emph{solver}, defined in Sec.~\ref{sec:data_generation}; (ii)~the fixed angles, obtained by an $\mathcal{O}(1)$ lookup into the existing database~\cite{WurtzGithub}; and (iii)~the deep-learning models, which predict angles in a single forward pass. 
For each method, we report the average per-instance cost of producing angles---the optimization time for the solver, the inference time for the learned models.
We exclude all one-time costs that amortize over all future instances, i.e., deep learning model trainings and precomputation of the fixed-angle database.

We average over the test set, defined in App.~\ref{app:training_data_split}, for Erd\H{o}s--R\'enyi and random-regular graphs, see Fig.~\ref{fig:pareto}(a) and~(b), and over the two heavy-hex graph families, see Fig.~\ref{fig:pareto}(c) and~(d).
The fixed-angle lookup and the solver span the two extremes of the frontier, see Fig.~\ref{fig:pareto}(a) and (c). 
A fixed-angle lookup is effectively free but yields the lowest ARs. 
By contrast, the solver achieves the highest quality at the cost of a full per-instance optimization, see e.g., Fig.~\ref{fig:pareto}(a).
The learned models sit between these extremes. 
In addition to the AR, we report the solver-relative AR defined as
\begin{equation}
    \mathrm{Reach}_i
    =
    \frac{\mathrm{AR}_{\mathrm{method},i}}
         {\mathrm{AR}_{\mathrm{solver},i}}
    \label{eq:reach}
\end{equation}
for an instance $i$.
Reach therefore directly measures how much of the \emph{solver} performance is recovered when the per-instance optimization is replaced by another angle-setting method.
It is an empirical normalization and not a theoretical approximation guarantee.

Importantly, the learned models are closer to the solver in quality and often closer to fixed angles in terms of runtime, see Fig.~\ref{fig:pareto}(b) and~(d). 
The model inference costs are several orders of magnitude below the \emph{solver}'s per-instance optimization time, while their approximation ratios generally recover over 99\% of the solver's accuracy. 
Because inference is a single forward pass whose cost is largely independent of the optimization landscape's ruggedness, whereas the solver's cost grows with depth, we see larger runtime gaps between learned models and the solver as $p$ increases. 

\subsection{Solver-relative performance across instances}
\label{sec:reach_distribution}

Since the mean Reach is consistently close to one across the learned models, average performance shows only small differences between the different model architectures. To capture these differences, we report the fraction of test instances that meet or exceed a given Reach threshold $\tau$.
For model $m$ and depth $p$, we define
\begin{equation}
    F_{m,p}(\tau)
    =
    \frac{1}{N}
    \sum_{i=1}^{N}
    \mathbb{I}
    \left[
        \mathrm{Reach}_{m,p,i} \geq \tau
    \right].
    \label{eq:reach_fraction}
\end{equation}
Here, $N$ is the size of the test set.

\begin{table*}[t]
\centering
\begin{minipage}[t]{0.6\textwidth}
    \centering
    \vspace{10pt}
    \setlength{\tabcolsep}{4pt}
    \renewcommand{\arraystretch}{1.2}
\begin{tabular}{lccccc}
    \hline\hline
    \textbf{Model}
    & \textbf{$p=1$}
    & \textbf{$p=2$}
    & \textbf{$p=3$}
    & \textbf{$p=4$}
    & \textbf{$p=5$} \\
    \midrule
    GNN
    & ${99.6 \pm 0.1}$
    & $\mathbf{95.1 \pm 0.2}$
    & $\mathbf{86.6 \pm 0.9}$
    & $\mathbf{84.4 \pm 0.8}$
    & $\mathbf{52.0 \pm 0.0}$ \\

    GIN
    & $\mathbf{99.7 \pm 0.1}$
    & $94.0 \pm 0.3$
    & $81.4 \pm 1.7$
    & $79.4 \pm 1.0$
    & $49.4 \pm 0.0$\\

    GT
    & $98.0 \pm 0.3$
    & $85.8 \pm 1.1$
    & $62.2 \pm 4.7$
    & $64.1 \pm 2.4$
    & $37.3 \pm 0.0$ \\

    GCN
    & $96.6 \pm 0.4$
    & $80.1 \pm 1.2$
    & $58.2 \pm 1.9$
    & $60.6 \pm 1.6$
    & $33.1 \pm 0.1$ \\

    DT
    & $95.2 \pm 0.3$
    & $78.5 \pm 0.5$
    & $57.5 \pm 0.5$
    & $60.7 \pm 1.0$
    & $39.9 \pm 0.0$ \\

    ET
    & $95.6 \pm 0.5$
    & $79.2 \pm 0.2$
    & $57.4 \pm 1.9$
    & $62.4 \pm 0.6$
    & $35.4 \pm 0.0$ \\

    MLP
    & $94.7 \pm 0.3$
    & $76.0 \pm 1.1$
    & $51.0 \pm 6.8$
    & $51.5 \pm 6.5$
    & $44.2 \pm 0.0$ \\
    \hline\hline
\end{tabular}
    \captionof{table}{
    Fraction of multi-topology test instances for which the predicted angles obtain at least $99.9\%$ of the solver AR. We report results across all depths $p \in \{1, \dots, 5\}$.
    Values are percentages and show mean $\pm$ standard deviation across five independent training runs.
    }
    \label{tab:reach_near_solver}
\end{minipage}
\hfill
\begin{minipage}[t]{0.35\textwidth}
    \vspace{0pt}
    \centering
    \includegraphics[width=0.7\linewidth]{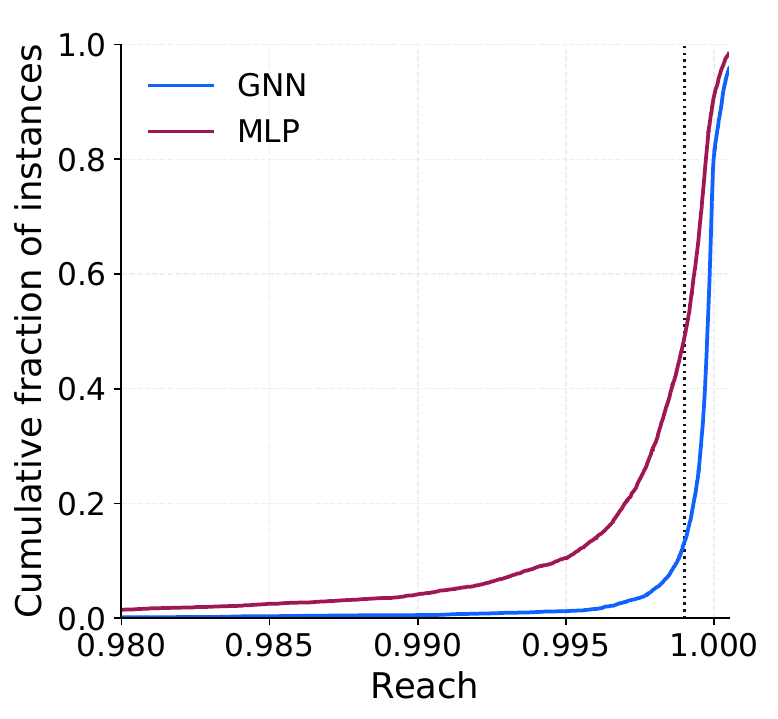}
    \captionof{figure}{
    Empirical cumulative distributions of instance-level Reach at $p=3$ (lower is better). Vertical line represents Reach of 99.9\% as an important threshold.
    }
    \label{fig:reach_distribution-main}
\end{minipage}
\end{table*}

We compute $F_{m,p}(\tau=99.9\%)$ separately for each training run and report the mean and standard deviation across the five random seeds.
The results reveal larger differences between the architectures than the mean Reach, see Tab.~\ref{tab:reach_near_solver}. 
Across the seven learned models, mean Reach ranges only from $99.973\%$ to $99.996\%$ at $p=1$ and from $93.025\%$ to $99.692\%$ at $p=5$. 
In contrast, the fraction of instances with $\mathrm{Reach}\geq99.9\%$ separates the models more clearly as circuit depth increases.
This difference is particularly pronounced for the GNN and GIN compared with the global-feature MLP. At $p=3$, the GNN obtains at least $99.9\%$ of the solver AR for $86.64\%$ of the test instances, compared with $51.02\%$ for the MLP. We visualize the strong differences in the corresponding cumulative distribution functions in Fig.~\ref{fig:reach_distribution-main}. 
At $p=4$, the corresponding fractions are $84.39\%$ and $51.48\%$.
The GIN shows a similar pattern, reaching the threshold for $81.40\%$ and $79.38\%$ of the instances at $p=3$ and $p=4$, respectively.

Overall, the GNN and GIN outperform the MLP more as $p$ increases.
This suggests that predicting QAOA angles based on graph structure is beneficial and that the models are not merely transferring angles based on concentration.

\subsection{Generalization to unseen graph families}

We now infer angles for 2D grid graphs and BA networks to study how our models generalize to unseen graph topologies. 
BA networks exhibit a scale-free power-law degree distribution characterized by heavy-tailed hubs: a small fraction of nodes maintain high degree while most have low degrees, see Fig.~\ref{fig:graph_topologies}(d). 
In contrast, 2D grids possess strong spatial locality with a bounded maximum degree of 4, see Fig.~\ref{fig:graph_topologies}(e). 
These graph families provide a clear structural contrast between scale-free connectivity and local spatial lattices.
They complement our training distribution, see Tab.~\ref{tab:graph_instances_dist}, thus allowing us to benchmark out-of-distribution performance. 
These topological differences cause a significant shift in $\beta$ angles found by the solver, see Fig.~\ref{fig:graph_topologies}(f).

\begin{figure}
    \centering
    \includegraphics[width=1\linewidth, clip, trim=10 0 12 0]{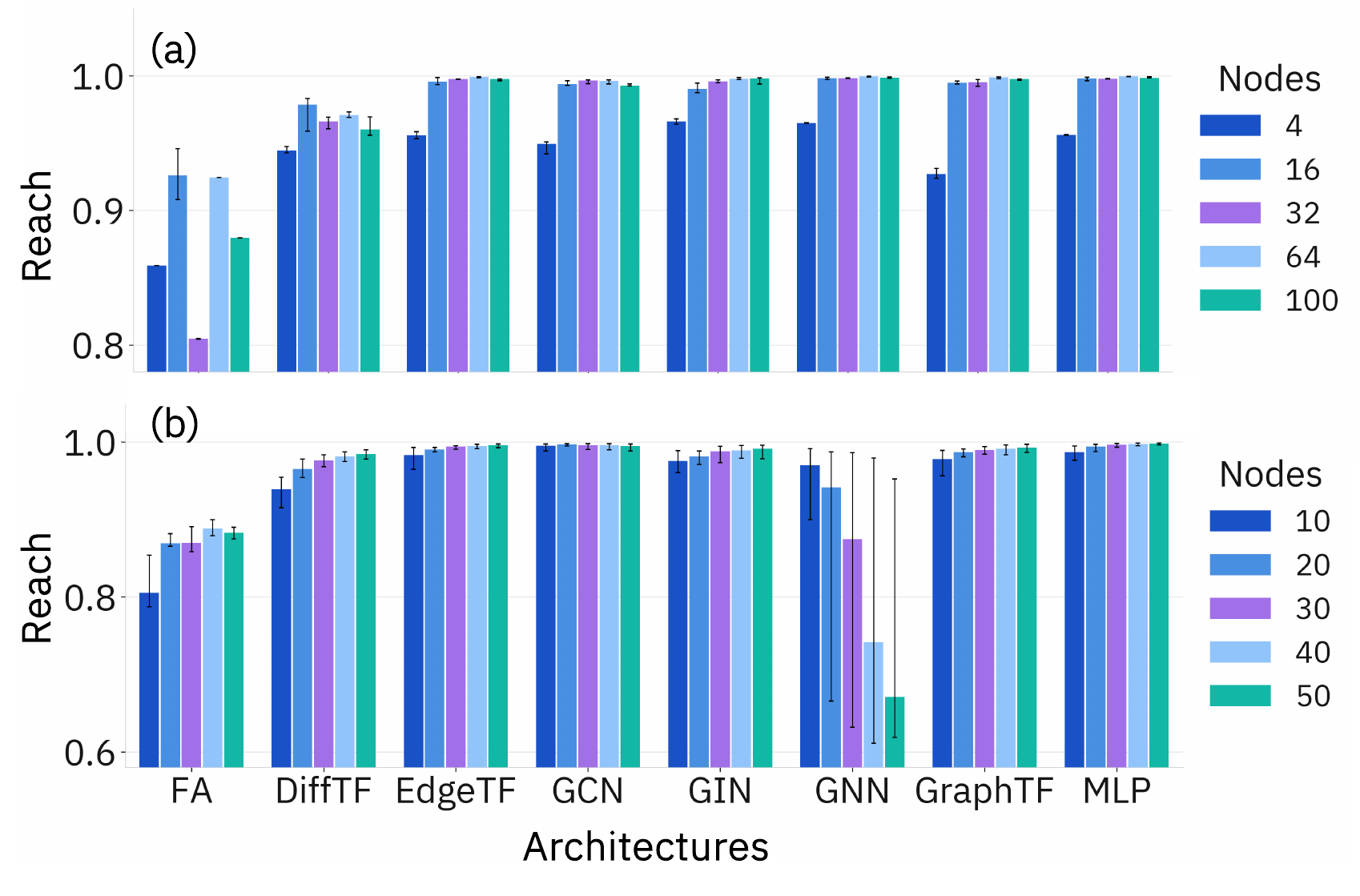}
   \caption{
   Inference results on unseen 2D grid (a) and Barabási-Albert (b) graph topologies for $p = 2$ at various node counts.
   The bars show the median Reach and the error bars correspond to the 25\% and 75\% quantiles across the five random seeds. 
   }
    \label{fig:ba_grid_combined}
\end{figure}

For both graph families, we evaluate the Reach for depth $p = 2$, the \emph{solver} found angles are discussed in App.~\ref{app:unseen_topology}.
As a baseline, we compare against the fixed-angles based on average node degree without any further optimization, see fixed-angles in Fig.~\ref{fig:ba_grid_combined}. 
Because the fixed angles are derived for unweighted regular graphs, applying them to our instances is an approximation.

Across both grid and Barabási–Albert (BA) graph topologies, the learned architectures consistently and substantially outperform the fixed-angle baseline, with median Reach improving from roughly 0.81–0.94 for fixed-angles to over 0.95 for most trained models. 
This improvement holds across nearly the full range of node counts tested (4–100 nodes for grids, 10–50 nodes for BA graphs), indicating that the learned parameter-prediction approach generalizes well beyond the naive heuristic baseline. 
Architectures such as EdgeTF, GCN, GIN, GraphTF, and MLP achieve high median Reach with tight, largely symmetric interquartile ranges throughout, reflecting both strong accuracy and low instance-to-instance variability as graph size increases. 
The one clear exception is the GNN architecture, which behaves very differently depending on graph topology: it performs reliably on grids but becomes unstable on BA graphs at higher node counts. The instability of the GNN is particularly interesting as GNN models outperform other models on in-distribution data. Thus, our results on BA graphs suggest that improved fitting of the structure of in-distribution graphs limits the transfer to unseen topologies. Contrary, generic models that leverage only aggregated data like MLP perform more reliably in transfer scenarios while they underperform on in-distribution data. 
Overall, the results show a strong zero-shot transfer from random regular, Erdős–Rényi, and heavy-hex training graphs to structurally distinct 2D grid and BA test graphs.

\subsection{Generalization to unseen graph sizes}

We now measure the performance of the models when applied to unseen graph sizes. 
We generate ten Erd\H{o}s--R\'enyi graphs with 150 and 200 nodes, and find their angles with the \emph{solver} for depths $p\in\{1, 2, 3\}$.
We then compare the inference results from our deep learning models to the performance of the solver and fixed angles. 
All deep learning models have only seen Erd\H{o}s--R\'enyi graphs with up to 100 nodes during training. 
Thus, all graphs in this experiment are outside of the training distribution with respect to node size. 
We provide an overview on how the training and testing distributions shift in this example in the appendix. 

\begin{figure}
    \centering
    \includegraphics[width=1\linewidth]{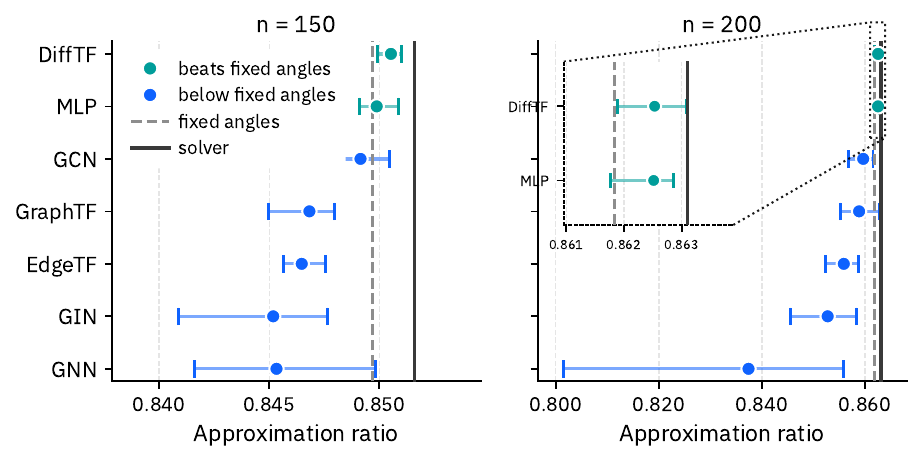}
    \caption{Inference on unseen graph sizes at the example of large-scale Erd\H{o}s--R\'enyi graph topologies with 150 nodes (blue) and 200 nodes (purple).}
    \label{fig:large-graphs}
\end{figure}

Overall, the angles inferred by graph models, e.g., GNN, GIN, under-perform plain transformer architectures, e.g., Edge transformer, Diffusion transformer, see Fig.~\ref{fig:large-graphs}. 
One possible explanation for this behavior is that message passing is better suited to capture specific graph characteristics and, therefore, fits the training data more strongly making generalization to unseen, larger graphs more challenging. 

\subsection{Limitations of MSE-style proxy objectives}

\begin{figure}[tbp]
    \centering
    \includegraphics[width=\linewidth]{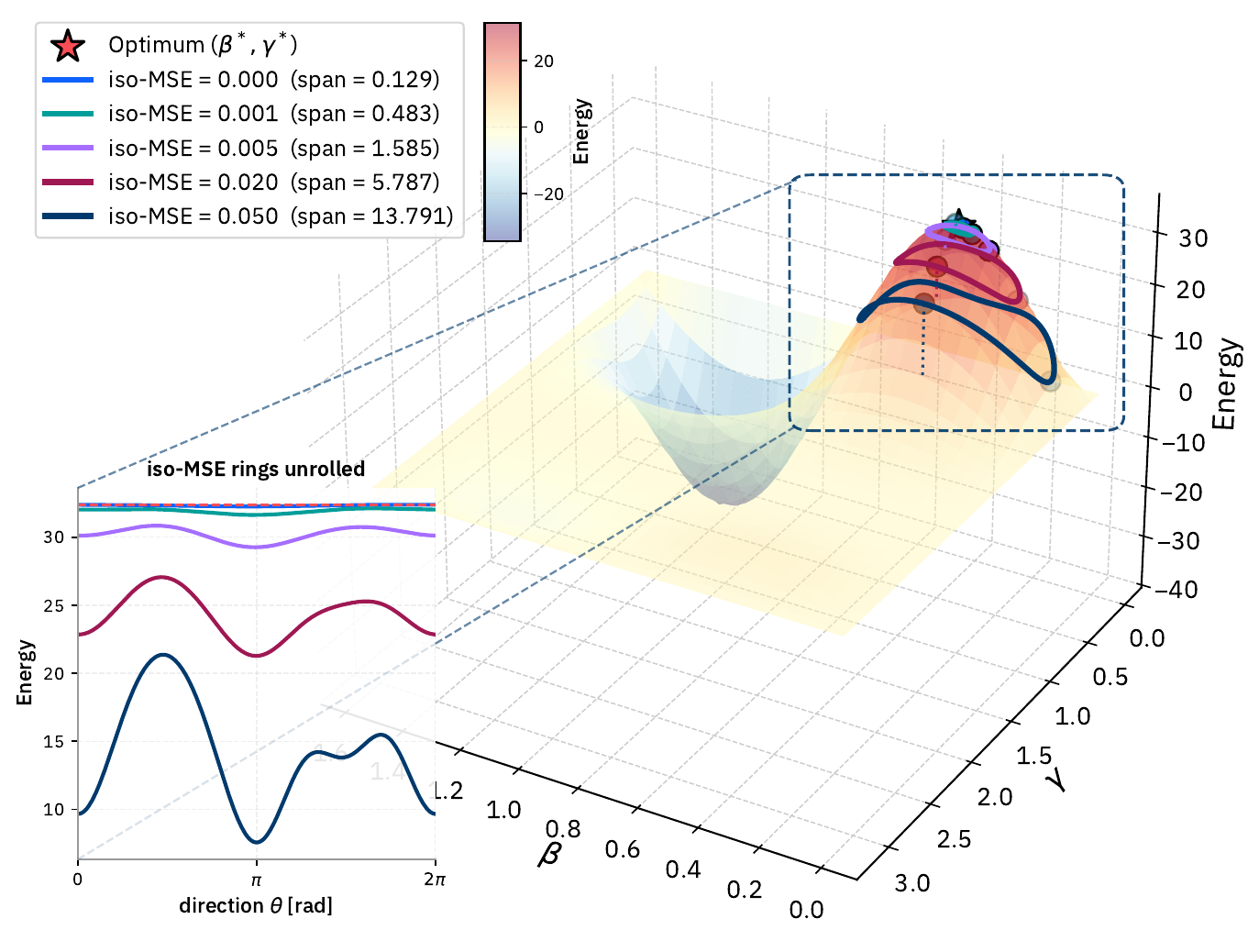}
    \caption{Visualizing de-correlation of an MSE proxy objective and the underlying energy landscape for increasing MSE errors at the example of a 100 node Erd\H{o}s--R\'enyi graph.}
    \label{fig:iso_mse_rings}
\end{figure}

Ideally, QAOA angles should minimize the energy $\langle H_C\rangle$.
However, the training objective for the models is the MSE with respect to pre-trained angles.
In the following ablation, we investigate potential de-correlation of our MSE training objective with $\langle H_C\rangle$. 
Conceptually, we visualize iso-MSE lines and the corresponding spans for an Erd\H{o}s--R\'enyi graph with 100 nodes and edge probability of $5\%$ from our validation dataset. 
In this example, the energy landscape is periodic, but as the distance to the optimum increases the iso-MSE rings become increasingly instable, see Fig.~\ref{fig:iso_mse_rings}.
Importantly, the span of the iso-MSE rings increase with increasing distance from the optimum, suggesting a decreasing correlation of the MSE objective with the energy landscape. 
Plotting the iso-MSE rings against the radiant shows increasing fluctuations when MSE scores decrease, see Fig.~\ref{fig:iso_mse_rings}. 
This suggests that larger MSE errors in the predicted angles correspond to less reliable energy estimates. 
Overall, for easy-to-fit periodic energy landscapes, MSE can serve as a proxy objective when MSE errors are low enough so that they are still reasonably well correlated to the energy landscape. 
However, for more complex energy landscapes, MSE objectives might be ill-defined for learning surrogates over the energy function. 

We highlight the de-correlation of standard deep learning regression losses and the underlying energy landscape because the important trade-off it implies: Deep learning regression objectives like the MSE are fast to compute, but can impose non-negligible inaccuracies.
More involved and more accurate objectives, like the energy function itself, imply significant computational overheads that slow down the training of deep learning models substantially---particularly with increasing QAOA depth. 

\subsection{Scaling behavior with dataset size}

\begin{figure}[tbp]
    \centering
    \includegraphics[width=1\linewidth]{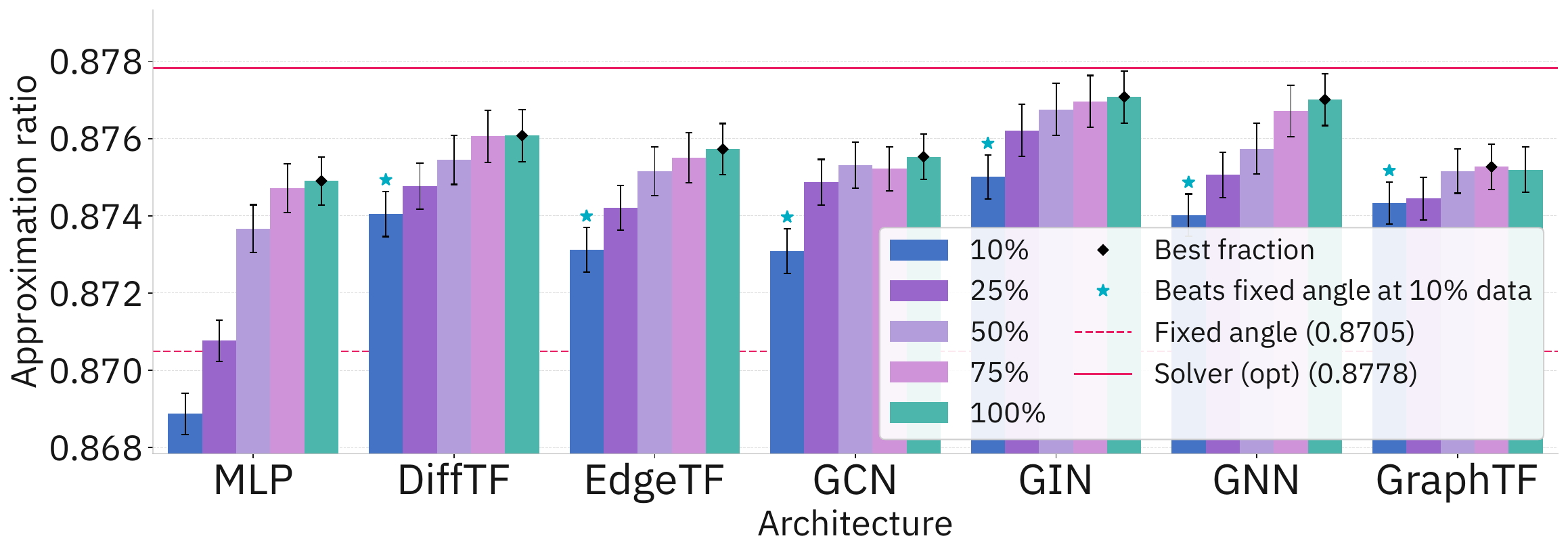}
    \caption{Inference across data fractions for all Deep Learning architectures at $p=3$. 
    These results demonstrate that even 10(\%) of the training data is sufficient to outperform fixed angles for all models except the MLP trained on 10(\%) of the training data.}
    \label{fig:data_scaling}
\end{figure}

To understand if we generate enough training data, or perhaps too much, we investigate model performance as a function of the dataset size. 
Overall, we observe a similar scaling between the different models of their ARs with increasing dataset size, see Fig.~\ref{fig:data_scaling} for $p=3$. 
The highest approximation ratios of 87.71\% and 87.70\% were observed from the GNN and GIN, respectively, with all of the training data. 
The MLP showed the lowest approximation ratio 86.89\% at 10$\%$ of the training data. 
Minor inconsistencies for GCN and Graph Transfomer are within the range of statistical noise. 
Importantly, all models that have access to the graph topology generally outperform the fixed-angles conjecture. 
This demonstrates that even with moderate data generation effort, data-driven approaches can outperform mathematically derived angles. 
With increasing dataset size, the models generally close the gap toward the \emph{solver}. 
This shows a clear correlation for the quality of the predicted angles and additional high quality data.

\subsection{Hardware validation}

\begin{figure}[htbp]
    \centering
    \includegraphics[width=\columnwidth]{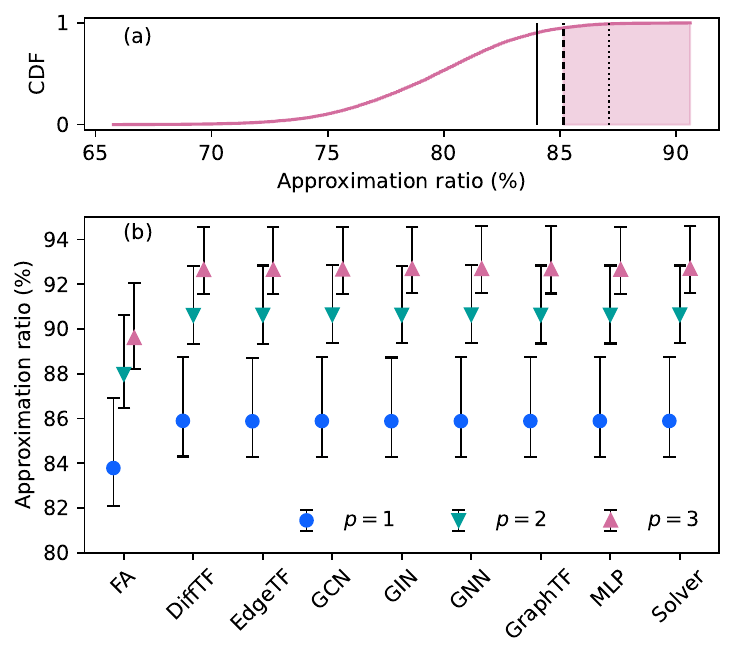}
    \caption{
    (a) Example $\mathrm{CVaR}_\eta$ computation.
    The pink line shows the cumulative distribution function (CDF) of the AR of the 8\,192 samples for a graph with 144 nodes.
    The solid, dashed, and dotted lines shows the $\eta=90\%$, $95\%$, and $99\%$ quantiles.
    The pink area represents the samples that we average over to compute $\mathrm{CVaR}_{95\%}$.
    (b) Average approximation ratio of 630 heavy-hex graphs measured on \emph{ibm\_boston}. 
    The lower error bars, markers, and upper error bars indicate the average over the 630 graphs of the $\mathrm{CVaR}_\alpha$ of the quantum samples at the $\alpha=90\%$, $95\%$, and $99\%$-level, respectively.
    FA indicates the fixed angles.}
    \label{fig:hw_validation}
\end{figure}

We now demonstrate the practical utility of our deep learning models by predicting angles for noisy quantum hardware. 
The training of the deep learning models and the evaluation results in previous sections are based on QAOA angles optimized by COBYLA with $\langle H_C\rangle$ evaluated using Pauli propagation.
Since this energy evaluation is approximate, we additionally validate the predicted QAOA angles on \emph{ibm\_boston}.
To avoid noise from SWAP routing, we validate 630 hardware-native heavy-hex graphs.
The graph sizes range from 12 to 144 nodes and, importantly, are not in the training data. 
For each deep learning model, we infer the QAOA angles and draw 8192 samples $x$ on \emph{ibm\_boston} from the corresponding QAOA circuit at depths $p=1$, $2$, and $3$.
The resulting sample distributions are compared to the \emph{solver} and the fixed angles.
As performance metric, we compute the Conditional Value at Risk at a level $\alpha$ ($\mathrm{CVaR}_\alpha$) of the ARs for each graph. 
We use CVaR since it is more robust to noise than the mean~\cite{Barron2024}. 
To compute the $\mathrm{CVaR}_\alpha$, we first calculate the AR of each sample $x_i$. 
Next, we average over the tail of the resulting distribution ranging from the $\alpha$ quantile to $1$, see an example for the $\alpha=95\%$ quantile in Fig.~\ref{fig:hw_validation}~(a).
For a given $\alpha$, each graph results in a $\mathrm{CVaR}_\alpha$ and we report the $\mathrm{CVaR}_\alpha$ averaged over all 630 graphs.
In Fig.~\ref{fig:hw_validation}~(b), we report CVaRs at $\alpha=90\%$, $\alpha=95\%$, and $\alpha=99\%$ for each method and observe that the performance of the proposed deep learning models is largely indistinguishable from the solver and outperforms the fixed-angle conjecture.
Therefore, for heavy-hex graphs, we simultaneously demonstrate both that our deep learning models show an excellent performance and Pauli propagation is a suitable approximate energy evaluator.

\section{Discussion and conclusion\label{sec:conclusion}}

In this work, we reframe QAOA angle selection as a supervised, graph-level regression problem, replacing a per-instance classical optimization with a single forward pass of a neural network. 
Trained on roughly twelve thousand weighted \textsc{MaxCut} instances across four graph families, a single model predicts angles that recover a large fraction of the solver's approximation ratio, while reducing the time to obtain angles by several orders of magnitude. 
A central ingredient that makes it possible to train a single model on the different graph families is the rescaling of the $\boldsymbol{\gamma}$ angles.
This rescaling lets one network learn across random regular, Erd\H{o}s--R\'enyi, and heavy-hex graphs at once.

Two aspects of our results are particularly relevant in practice. 
First, the learned models sit in a favorable part of the runtime--accuracy Pareto fronts. 
They are close to the solver in quality and approach the fixed-angle lookup in runtime.
Because inference is a single forward pass, its cost is largely independent of circuit depth, so the runtime advantage over the \emph{solver} grows as $p$ increases. 
The instance-level analysis further shows that similar mean performance can mask differences in consistency across graphs. 
On the multi-topology test set, the more advanced graph models (GNN, GIN) most consistently achieve near-solver performance, while the MLP remains competitive in mean Reach but performs less consistently across individual instances as circuit depth increases.
Second, a single trained model can transfer to unseen graphs without retraining or fine-tuning. 
Some models generalize to Erd\H{o}s--R\'enyi graphs of $150$ and $200$ nodes, and to Barab\'asi--Albert and 2D-grid topologies not seen during training. 
Here, we observe a consistent pattern: plain transformer architectures generalize to larger and unseen graphs more reliably than message-passing networks, which appear to fit the training topologies more strongly at the expense of transfer. 
The strong distribution shift we observe for 2D grid graphs marks the point at which topology-specific behavior begins to dominate.

Our analysis also clarifies the limits of the approach. The MSE objective is only a proxy for the energy, and holds only where the two remain correlated. 
As our iso-MSE analysis shows, this correlation degrades with increasing distance from the optimum, so that a given MSE error translates less reliably into an energy error on more rugged landscapes. 
This points to a trade-off between fast, differentiable regression losses and slower but more accurate energy-based objectives, which becomes more pressing at higher depth. 
A related limitation is that our labels inherit the approximations of the data-generation pipeline, in particular the truncated Pauli-propagation evaluation for $p>1$, which bounds the quality any model trained against them can reach.

These limitations suggest several directions for future work. 
Closing the remaining gap to the solver may require moving beyond pure regression towards objectives that incorporate the energy more directly, or using the predicted angles as warm starts for a short classical refinement that recovers the final percentage points of AR at little additional cost. 
Extending the formulation beyond \textsc{MaxCut} to other problems, and validating the predicted angles on quantum hardware rather than in simulation, are natural next steps towards practical deployment.

\section{Data Availability}

We open-source our trained models and inference code across depths $p \in \{1,\dots,5\}$, providing a fast, drop-in replacement for per-instance QAOA optimization that generalizes across graph sizes and topologies. The code we used to create and train the deep learning models can be found under \url{https://github.com/jhnnsjkbk1/qaoa_ai_model_training}. We open-source pretrained models at \url{https://huggingface.co/collections/ibm-research/angles-for-quantum-approximate-optimization}, which can be used via an inference engine provided by the QAOA Training Pipeline \url{https://github.com/qiskit-community/qaoa_training_pipeline}. More broadly, we hope that reframing angle selection---from optimizing each instance separately to learning the map from problem structure to angles---helps remove one of the standing bottlenecks to deploying QAOA at scale.

\section{Acknowledgments}

We are thankful for the help of Romeo Kienzler and Daniela Szwarcman from IBM Research that allowed to enhance the code base and providing ONNX support for the deep learning models.
We are thankful to Ehud Karavani and Yishai Shimoni for providing us with fixed-angles for the heavy-hex graphs.
We acknowledge Stefan Woerner for useful conversations.

\bibliography{refs_clean}

\appendix

\section{Experimental settings of prior work}
\label{app:rw_settings}

Research on ML approaches for setting QAOA angles is difficult to compare, as methods are evaluated using different problem sizes, circuit depths, and dataset sizes.
Furthermore, some approaches apply additional classical optimization and performance results can be obtained with simulation or quantum hardware. 
Table~\ref{tab:qrw_settings} summarizes these evaluation choices for the methods discussed in Sec.~\ref{sec:related-work}, grouped according to the introduced taxonomy. 
Four patterns are relevant to our work. 

First, direct angle prediction is largely studied on small graphs containing up to 20 nodes, with fixed-depth experiments reaching at most $p=4$, with the exception of QAOA-GPT that uses ADAPT-QAOA. 
Our models instead are trained on graphs with up to 144 nodes and evaluated on graphs with up to 200 nodes and depths up to $p=5$. Prior angle transfer methods evaluate larger target graphs, most notably 100 nodes at $p=3$ in \citet{falla2024graph}, but reuse angles obtained from other instances.

Second, some works evaluate ML-predicted angles directly, while others further optimize them. 
For instance, \citet{meng2025conditional} apply 100 optimization steps after ML prediction, QSeer~\cite{jiang2025qseer} reports both direct and 100-step refined results, and prior depth-extrapolation approaches similarly use their predictions to initialize further optimization. 
Such refined results are therefore not directly comparable to direct prediction methods. 
Our models instead predict the full QAOA angle vector without instance-specific post-prediction optimization.

Third, most prior methods are evaluated using classical quantum-circuit simulation, while hardware evaluation remains uncommon. 
DARBO~\cite{cheng2024quantum} reports both simulation and experiments on superconducting quantum hardware, whereas RL-RQAOA~\cite{patel2024reinforcement} computes the required depth-one QAOA quantities classically using analytical expressions. 
Alongside our simulation-based results, we also evaluate our predictions on IBM superconducting quantum hardware for heavy-hex graphs up to $p=3$.

Finally, public code and pretrained models are available only for a subset of methods. Neural QAOA$^2$~\cite{Zheng2026} and \citet{xie2023quantum} provide both, while \citet{amosy2024iteration}, \citet{falla2024graph}, and \citet{wauters2020reinforcement} release code only. We release training and inference code together with pretrained model weights.

\begin{table*}[t]
\centering

\setlength{\tabcolsep}{2.5pt}
\renewcommand{\arraystretch}{1.5}
\fontsize{7.7}{8.7}\selectfont

\caption{Comparison of ML methods for QAOA angle setting}
\label{tab:qrw_settings}

\begin{NiceTabular}{
L{2.80cm}
L{1.80cm}
C{1.50cm}
C{0.95cm}
L{1.75cm}
L{3.15cm}
L{1.85cm}
L{1.2cm}
L{1.40cm}
}
\hline\hline
\textbf{Method} &
\textbf{Problem} &
\textbf{Size $n$} &
\textbf{$p_{\max}$} &
\textbf{Training} &
\textbf{Instances} &
\textbf{Post-opt.} &
\textbf{Eval.} &
\textbf{Release}\\
\midrule

\multicolumn{9}{@{}l}{\textbf{Direct angle prediction}}\\

\citet{amosy2024iteration} &
MC (uw) &
$16$ &
$2$ ($4$) &
$5$k/setup &
ER &
$0$ / $\sim1$ iter. &
Sim. &
Code \\

\citet{deshpande2022capturing} &
MC (uw) &
$8$ ($9$) &
$3$ &
$12$k\derived &
connected, non-isomorphic &
-- &
Sim. &
-- \\

\citet{liang2024graph} &
MC (uw) &
$15$ &
$1$\derived &
$9.5$k \derived &
RR &
-- &
Sim. &
-- \\

QSeer~\cite{jiang2025qseer} &
MC (w/uw) &
$20$ &
$3$ &
$>200$k\derived &
non-isomorphic, ER (OOD: RR) &
$0$ / $100$ iter. &
Sim. &
-- \\

\citet{meng2025conditional} &
MC (uw) &
$8$ ($16$) &
$3$ &
$10$k &
random, RR, WS &
$100$ iter. &
Sim. &
-- \\

QAOA-GPT~\cite{Tyagin2025} &
MC (w) &
$14$ &
adapt. &
$>50$k/setup &
ER (OOD: BA, bipartite, RR, WS) &
-- &
Sim. &
-- \\

\rowcolor{blue!8}
\textbf{Ours} &
MC (w/uw) &
$144$ ($200$) &
$5$ &
$12$k &
RR, ER, heavy-hex (OOD: BA, Grid2D) &
-- &
Sim. + HW &
Code + model \\

\addlinespace[2pt]
\midrule
\addlinespace[1pt]
\multicolumn{9}{@{}l}{\textbf{Angle transfer}}\\

\citet{moussa2022unsupervised} &
MC (uw); dense QUBO &
$18$ &
$3$ &
$240$ \derived &
ER; dense &
0 ($\leq10$ calls) &
Sim. &
-- \\

GAT-MIS~\cite{xu2025qaoa} &
MIS (uw) &
$14$ ($25$) local &
$2$ &
$3$k &
ER (OOD: SuiteSparse) &
-- &
Sim.&
-- \\

\citet{falla2024graph} &
MC (uw) &
$40$ ($100$) \newline
$20$, $p=10$ &
$10$ &
$30$k ($p=3$);\newline
$4.5$k ($p=10$) &
random (OOD: RR, WS) &
$0$ / $10$ iter. &
Sim. &
Code \\

\addlinespace[2pt]
\midrule
\addlinespace[1pt]
\multicolumn{9}{@{}l}{\textbf{Depth extrapolation}}\\

\citet{alam2020accelerating} &
MC (uw) &
$8$ &
$6$ &
$66$ &
ER &
to $10^{-6}$ tol. &
Sim. &
-- \\

\citet{xie2023quantum} &
MC (uw) &
$8$ &
$10$ &
$66$ &
ER &
$34-1.2$k calls &
Sim. &
Code + model \\

\addlinespace[2pt]
\midrule
\addlinespace[1pt]
\multicolumn{9}{@{}l}{\textbf{Learned optimization}}\\

\citet{khairy2020learning} &
MC (uw) &
$8$ ($22$) &
$4$ &
$7$ &
ER, ladder, barbell, caveman &
$0$ / $96$ calls &
Shot sim. &
-- \\

PG-QAOA~\cite{yao2020policy} &
Quantum control &
$20$ &
$60$ &
Per-instance &
Ising chains &
\NA &
Sim. &
-- \\

\citet{wauters2020reinforcement} &
TFIM &
$8$ ($128$) &
$15$ &
$1$/size &
uniform / disordered chains &
opt. (iter. \NR) &
Sim. &
Code \\

RL-RQAOA~\cite{patel2024reinforcement} &
MC (w) &
$200$ &
$1$ &
Per-instance &
RR, 3-RR, cage &
\NA &
Analytical &
Partial code \\

\citet{chandarana2023meta} &
MC (w/uw); SK &
$10$ ($18$) &
$2$ &
$600$\derived &
3-RR, complete &
$\leq200$ iter. &
Sim. &
-- \\

\citet{chen2025learning} &
MC; SK &
$7$ ($16$)&
$1$\derived &
\NR &
ER, complete &
\NA &
Sim. &
-- \\

DARBO~\cite{cheng2024quantum} &
MC (w); QUBO &
$16$ ($5$ HW) &
$10$ &
Per-instance &
3-RR; general QUBO &
\NA &
Sim. + HW &
Code \\

Neural QAOA$^2$~\cite{Zheng2026} &
MC (w); QUBO; Ising &
$10$ local &
$3$ &
$200$\derived &
benchmark graphs &
$20$ iter. &
Sim. &
Code + model \\

SSR-MCTS~\cite{agirre2025monte} &
MC (uw); MAX-3-SAT &
$11$ &
$10$ &
Per-instance &
3-RR (MC); -- (3-SAT) &
opt. (iter. \NR) &
Sim. &
Code \\

QAOA-Predictor~\cite{Coelho2026} &
12 QUBO classes &
$15$ ($20$) &
$500$ &
$92$k &
ER, BA, complete bipartite &
-- &
Shot sim. &
-- \\

\hline\hline
\end{NiceTabular}

\vspace{3pt}

\begin{minipage}{\textwidth}
\fontsize{8.4}{9.5}\selectfont
\RaggedRight

\textbf{Size.}
Refers to the largest training graph size, with larger evaluation sizes in parentheses, for per-instance methods, it gives the largest evaluated graph. \emph{Local} indicates that the method evaluates global graphs through local subproblems.

\textbf{Training.}
Counts refer to unique graph/problem instances, excluding repeated depths,
angle sets, labels, configurations, and episodes; \emph{per-instance}
denotes no cross-instance training.

\textbf{Post-opt.}
-- denotes no further optimization after ML angle setting; \emph{opt.} denotes optional optimization.

\textbf{Public code/models.}
\emph{Code + model} denotes publicly available implementation and pretrained model weights; \emph{code} denotes a public implementation without pretrained weights; \emph{partial code} denotes an incomplete training or inference workflow.

\textbf{Notation.}
MC: MaxCut; w/uw: weighted/unweighted; HW: Hardware; ER: Erd\H{o}s--R\'enyi;
RR: random regular; WS: Watts--Strogatz; BA: Barab\'asi--Albert;
TFIM: transverse-field Ising model; \derived{} inferred from the reported
setup; \NR: not reported; \NA: not applicable.
\end{minipage}

\end{table*}

\section{Dataset generation and validation}
\label{app:dataset_details}

This section provides details on the dataset and training-label generation described in Sec.~\ref{sec:data_generation}. 
We analyze QAOA performance and detail the data split design choices for the model training.

\subsection{Distribution of QAOA angles and ARs}
\label{app:training_distributions}

\begin{figure}
    \centering
    \includegraphics[width=1\linewidth]{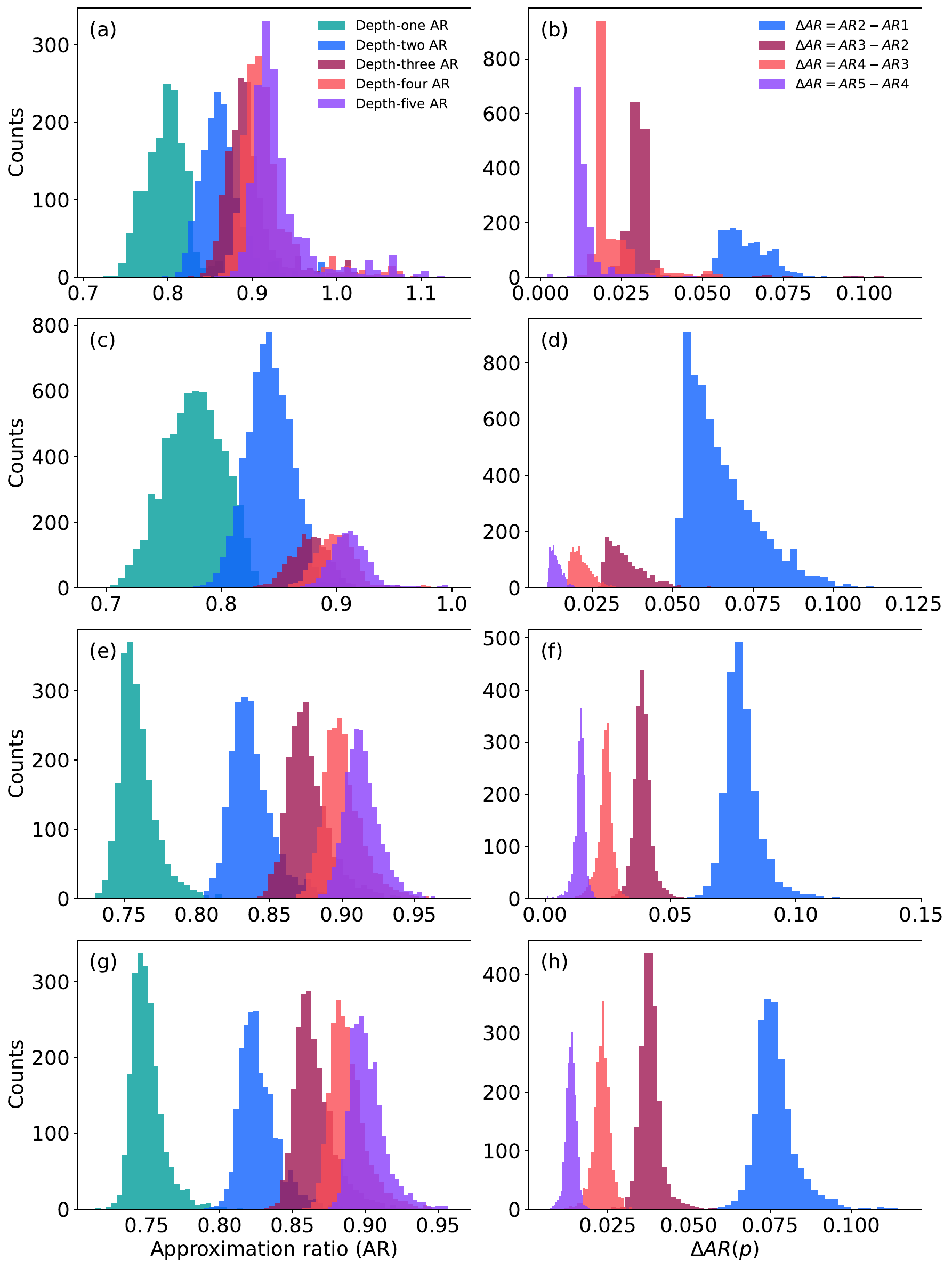}
    \caption{Distribution of AR, and the delta in per-instance Ars between successive circuit depth. 
    (a) and (b) AR and $\Delta\mathrm{AR}$ for Random d-regular graphs.
    (c) and (d) AR and $\Delta\mathrm{AR}$ for Erd\H{o}s--R\'enyi graphs.
    (e) and (f) AR and $\Delta\mathrm{AR}$ for heavy-hex graphs with uniform edge weights.
    (g) and (f) AR and $\Delta\mathrm{AR}$ for heavy-hex graphs with discrete edge weights.}
    \label{fig:AR_and_deltas_all_graphs}
\end{figure}

Since model quality depends on the quality of the training data we carefully validate that the QAOA angles produce high AR increasing with QAOA depth $p$ for all graph instances.
For each graph the \emph{solver} gives us the optimized angles and the corresponding expected energy, evaluated with Pauli propagation.
We convert this energy to a cut value and then to an AR.
We plot the distribution of the ARs for the different graph families in Fig.~\ref{fig:AR_and_deltas_all_graphs} and observe high ARs that increase with $p$.
A few graph instances show an $\mathrm{AR}>1$ which we attribute to approximation errors in the QAOA energy introduced by Pauli propagation.
Furthermore, for each graph instance we compute the difference in AR between $p$ and $p+1$ which we label $\Delta\mathrm{AR}$.
All but ten instances have $\Delta\mathrm{AR}>0$ indicating that the QAOA at depth $p+1$ has a smaller Pauli propagation estimated energy than the QAOA at depth $p$, as expected.

\subsection{Training, validation, and test data split}
\label{app:training_data_split}

To construct representative datasets for training and evaluation, we first split the labeled problem instances within each graph family into disjoint training, validation, and test sets and then combine the corresponding subsets across graph families. 
The training set is used to update the model parameters, while the validation set is evaluated during training and used exclusively for early stopping when the validation loss no longer improves. 
The test set is held out, with graph instances kept consistent across circuit depth, from both model training and model selection and is used only for the final reported results.
Within each graph family, 15\% of the instances are first reserved for testing. 
Of the remaining instances, 90\% are assigned to the training set and 10\% to the validation set, see Tab.~\ref{tab:sample_counts} for the resulting counts. 
The splits are constructed to approximately preserve the distributions of graph size and topology-specific parameters, such as degree for random $d$-regular graphs and edge probability for Erd\H{o}s--R\'enyi graphs, see Tab.~\ref{tab:graph_instances_dist}. 

While we aim to maintain comparable dataset and split sizes across circuit depths, some differences remain. 
First, the number of Erd\H{o}s--R\'enyi instances is reduced at higher depths to limit computational cost during label generation and validation, see Tab.~\ref{tab:sample_counts}. 
Second, several of the remaining large and dense Erd\H{o}s--R\'enyi and random $d$-regular graphs at higher depths could not be optimized within the available compute budget and were therefore excluded. 
Third, the ten heavy-hex instances at $p=5$ with a lower energy than at $p=4$ were removed since it indicates an unsuccessful optimization.

\begin{table}[tbp]   
    \centering
    \caption{Number of instances in the training, validation, and test sets for each graph family and QAOA depth. The union across graph families forms the corresponding multi-topology datasets.}
    \label{tab:sample_counts}

    \begin{tabular*}{\columnwidth}{@{\extracolsep{\fill}}llrrrr@{}}
        \hline\hline
        Depth & Graph family & Train & Validation & Test & Total \\
        \hline
        \multirow{4}{*}{$p=1$}
            & Erd\H{o}s--R\'enyi     & 4{,}896& 544 & 216 &5{,}656\\
            & Random Regular         & 1{,}194& 133 & 227& 1{,}554 \\
            & Heavy Hex (Uniform)    & 1{,}606 & 179 & 315 & 2{,}100 \\
            & Heavy Hex (Discrete)   & 1{,}606 & 179 & 315 & 2{,}100 \\
        \addlinespace
        \multirow{4}{*}{$p=2$}
            & Erd\H{o}s--R\'enyi     & 4{,}896& 544 & 216 & 5{,}656\\
            & Random Regular         & 1{,}194& 133 & 227& 1{,}554 \\
            & Heavy Hex (Uniform)    & 1{,}606 & 179 & 315 & 2{,}100 \\
            & Heavy Hex (Discrete)   & 1{,}606 & 179 & 315 & 2{,}100 \\
        \addlinespace
        \multirow{4}{*}{$p=3$}
            & Erd\H{o}s--R\'enyi     & 1{,}237& 125& 216& 1{,}578 \\
            & Random Regular         & 1{,}194& 133 & 227& 1{,}554 \\
            & Heavy Hex (Uniform)    & 1{,}606 & 179 & 315 & 2{,}100 \\
            & Heavy Hex (Discrete)   & 1{,}606 & 179 & 315 & 2{,}100 \\
        \addlinespace
        \multirow{4}{*}{$p=4$}
            & Erd\H{o}s--R\'enyi     & 1{,}221& 125& 216& 1{,}562 \\
            & Random Regular         & 1{,}194& 133 & 227& 1{,}554 \\
            & Heavy Hex (Uniform)    & 1{,}606 & 179 & 315 & 2{,}100 \\
            & Heavy Hex (Discrete)   & 1{,}606 & 179 & 315 & 2{,}100 \\
        \addlinespace
        \multirow{4}{*}{$p=5$}
            & Erd\H{o}s--R\'enyi     &  1{,}137&  118&  216&  1{,}471\\
            & Random Regular         &  1{,}163&  131&  227&  1{,}521\\
            & Heavy Hex (Uniform)    & 1{,}602 & 175 & 315 & 2{,}092 \\
            & Heavy Hex (Discrete)   & 1{,}605 & 178 & 315 & 2{,}098 \\
        \hline\hline
    \end{tabular*}
\end{table}

\section{Additional results and ablations}

This section provides additional analyses that complement the main results.
We first examine the distribution of solver-relative performance (Reach) and then study selected modeling choices through targeted ablations.

\subsection{Solver-relative performance distributions}
\label{app:reach_distribution}

Here, we examine the instance-level Reach distributions and additional lower-tail statistics in more detail.
Besides the mean Reach, we report the lower fifth and tenth percentiles and the minimum Reach for all models, see Tab.~\ref{tab:reach_distribution_full}.
These quantities characterize the lower-performing part of the instance distribution without relying on a single threshold. E.g., for $p=5$, the 5\% instances on which the GNN performs worst still obtain 97.47\% of the solver accuracy on average.
For each training run, the statistics are first computed across test instances and are then averaged across the five random seeds. 
We additionally report the lowest minimum Reach observed across any of the five runs. For example, for the GNN model on $p=5$, the instance that performs worst across all five seeds still attains 93.41\% of the solver performance. 
The lower-tail statistics show the same pattern as the cumulative distributions. 
However, differences are more visible in the lower end of the distribution.
At $p=3$, for example, the fifth percentile is $99.80\%$ for the GNN and $99.13\%$ for the MLP.
At $p=4$, the corresponding values are $99.76\%$ and $99.01\%$. 
The minimum values show larger instance-specific deviations, while the fifth and tenth percentiles remain substantially closer to the \emph{solver}.
The GNN and GIN provide the strongest overall lower-tail performance at the larger circuit depths.


\begin{table*}[t]
\centering
\caption{
Distribution of solver-relative Reach on the multi-topology test set.
All values are percentages.
Mean Reach, the minimum, and the fifth and tenth percentiles are averaged across five independent training runs.
The worst minimum is the lowest instance-level Reach observed across the five runs.
$F_{0.999}$ denotes the fraction of test instances with $\mathrm{Reach}\geq0.999$ and is reported as mean $\pm$ standard deviation across seeds.
}
\label{tab:reach_distribution_full}
\setlength{\tabcolsep}{3.6pt}
\renewcommand{\arraystretch}{1.08}
\begin{tabular}{c l r r r r r r}
\hline\hline
$\mathbf{p}$
& \textbf{Model}
& \textbf{Mean}
& \textbf{Min.}
& \textbf{Worst min.}
& \textbf{5th perc.}
& \textbf{10th perc.}
& $\mathbf{F_{0.999}}$ \\
\midrule

1 & GNN     & 99.996 & 99.76 & 99.74 & 99.99 & 99.99 & $99.62 \pm 0.05$ \\
1 & GIN     & 99.994 & 99.67 & 99.58 & 99.98 & 99.99 & $99.65 \pm 0.07$ \\
1 & GraphTF & 99.988 & 99.40 & 99.22 & 99.95 & 99.97 & $98.02 \pm 0.27$ \\
1 & GCN     & 99.981 & 99.43 & 99.37 & 99.93 & 99.96 & $96.64 \pm 0.39$ \\
1 & DiffTF  & 99.977 & 99.17 & 99.02 & 99.90 & 99.95 & $95.23 \pm 0.32$ \\
1 & EdgeTF  & 99.974 & 99.20 & 99.13 & 99.91 & 99.94 & $95.55 \pm 0.45$ \\
1 & MLP     & 99.973 & 99.17 & 99.12 & 99.90 & 99.94 & $94.70 \pm 0.31$ \\

\midrule
2 & GNN     & 99.983 & 98.41 & 98.38 & 99.90 & 99.95 & $95.09 \pm 0.19$ \\
2 & GIN     & 99.976 & 98.31 & 98.20 & 99.89 & 99.93 & $93.99 \pm 0.30$ \\
2 & GraphTF & 99.949 & 98.29 & 98.17 & 99.79 & 99.87 & $85.78 \pm 1.07$ \\
2 & GCN     & 99.932 & 98.47 & 97.93 & 99.74 & 99.83 & $80.12 \pm 1.19$ \\
2 & DiffTF  & 99.928 & 98.12 & 98.06 & 99.69 & 99.80 & $78.51 \pm 0.54$ \\
2 & EdgeTF  & 99.929 & 98.07 & 97.99 & 99.70 & 99.80 & $79.23 \pm 0.23$ \\
2 & MLP     & 99.914 & 97.89 & 97.76 & 99.66 & 99.78 & $75.96 \pm 1.06$ \\

\midrule
3 & GNN     & 99.946 & 94.54 & 94.06 & 99.80 & 99.88 & $86.64 \pm 0.88$ \\
3 & GIN     & 99.930 & 94.32 & 93.41 & 99.74 & 99.83 & $81.40 \pm 1.72$ \\
3 & GraphTF & 99.830 & 94.25 & 93.91 & 99.34 & 99.62 & $62.21 \pm 4.72$ \\
3 & GCN     & 99.769 & 89.07 & 87.53 & 99.31 & 99.60 & $58.20 \pm 1.91$ \\
3 & DiffTF  & 99.820 & 94.59 & 94.31 & 99.35 & 99.58 & $57.50 \pm 0.52$ \\
3 & EdgeTF  & 99.814 & 94.46 & 94.35 & 99.35 & 99.59 & $57.39 \pm 1.92$ \\
3 & MLP     & 99.759 & 93.88 & 92.54 & 99.13 & 99.47 & $51.02 \pm 6.81$ \\

\midrule
4 & GNN     & 99.934 & 97.28 & 96.55 & 99.76 & 99.86 & $84.39 \pm 0.81$ \\
4 & GIN     & 99.921 & 97.65 & 96.12 & 99.70 & 99.82 & $79.38 \pm 0.95$ \\
4 & GraphTF & 99.803 & 92.11 & 88.30 & 99.32 & 99.64 & $64.06 \pm 2.36$ \\
4 & GCN     & 99.742 & 87.58 & 82.80 & 99.26 & 99.61 & $60.62 \pm 1.59$ \\
4 & DiffTF  & 99.834 & 93.63 & 93.40 & 99.43 & 99.64 & $60.67 \pm 0.98$ \\
4 & EdgeTF  & 99.837 & 93.38 & 93.17 & 99.41 & 99.63 & $62.37 \pm 0.59$ \\
4 & MLP     & 99.727 & 92.04 & 88.77 & 99.01 & 99.46 & $51.48 \pm 6.54$ \\

\midrule
5 & GNN     & 99.692 & 88.79 & 86.35 & 98.91 & 99.31 & $52.02\pm 0.01$ \\
5 & GIN     & 99.544 & 88.14 & 69.04 & 97.95 & 99.05 & $49.43 \pm 0.01$ \\
5 & GraphTF & 98.461 & 77.31 & 50.43 & 93.94 & 95.37 & $37.35 \pm 0.03$ \\
5 & GCN     & 93.025 & 54.39 & 47.84 & 66.57 & 71.10 & $33.14 \pm 0.12$ \\
5 & DiffTF  & 98.073 & 78.86 & 76.35 & 91.91 & 93.55 & $39.90 \pm 0.03$ \\
5 & EdgeTF  & 99.260 & 87.22 & 87.16 & 97.20 & 97.96 & $35.47 \pm 0.01$ \\
5 & MLP     & 99.606 & 89.68 & 87.45 & 98.45 & 99.25 & $44.28 \pm 0.01$ \\

\hline\hline
\end{tabular}
\end{table*}

\subsection{Joint versus topology-specific training}
\label{app:joint_training}

\begin{figure}[htbp]
    \centering
    \includegraphics[width=\columnwidth]{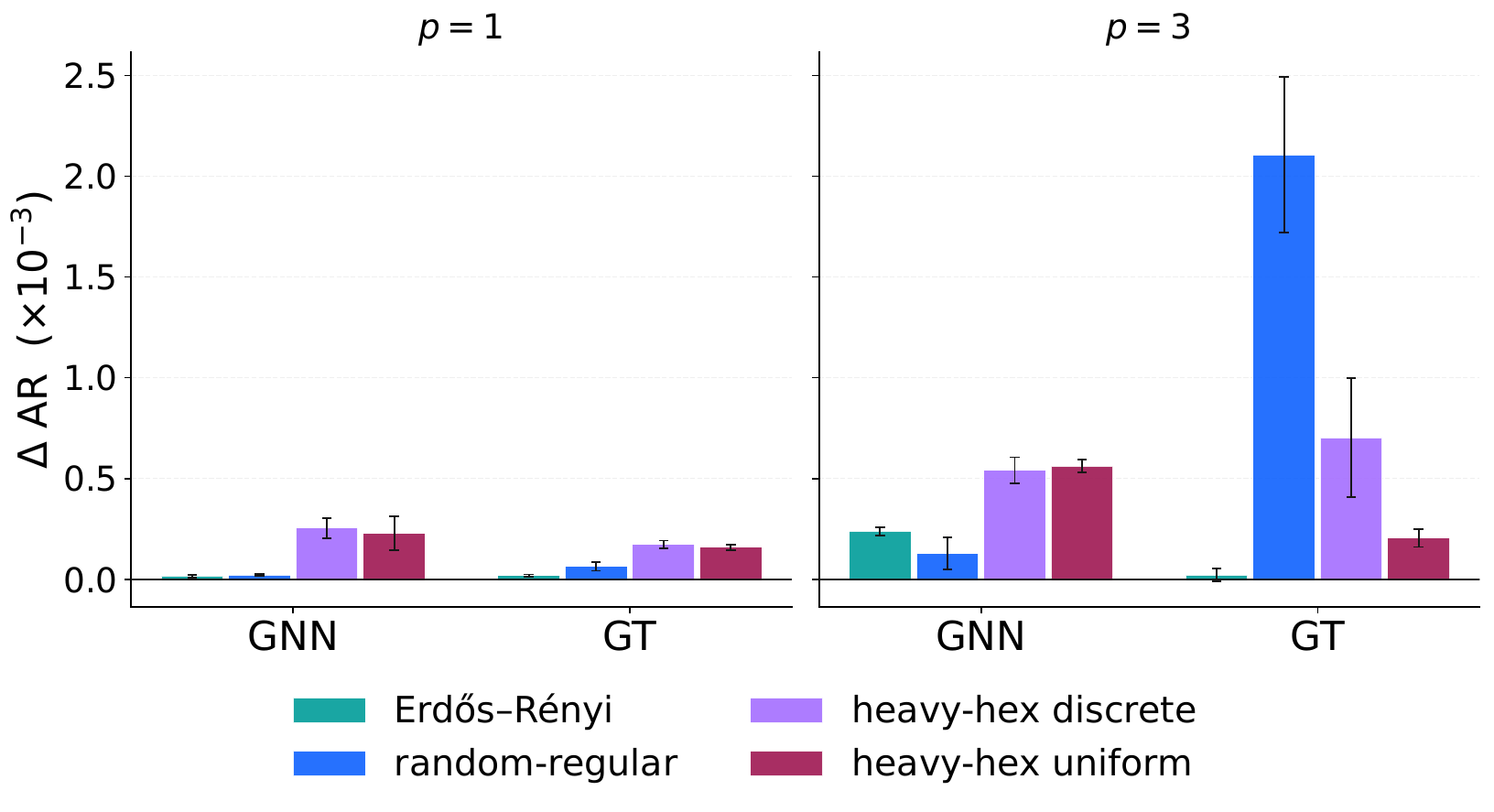}
    \caption{
    Difference in approximation ratio between models trained jointly across all graph families and topology-specific models at $p=1$ and $p=3$.
    Results are shown for the GNN and Graph Transformer across the four training graph families.
    Positive values indicate higher approximation ratio for joint training.
    Error bars show variation across the five independent training runs.
    }
    \label{fig:joint_vs_specialist}
\end{figure}

Our main models are trained jointly across all graph families.
To assess whether joint training reduces predictive performance, we additionally train the GNN and the Graph Transformer models separately on each graph family.
Overall, joint training closely matches topology-specific training.
The differences in AR are small across graph families and both architectures, see Fig.~\ref{fig:joint_vs_specialist} for $p=1$ and $p=3$.
At $p=1$, all differences are below $3\times10^{-4}$.
At $p=3$, the largest difference is about $2\times10^{-3}$ and occurs for the Graph Transformer on random $d$-regular graphs, where joint training performs better.
These results show that a single model can be trained across multiple graph families without a consistent loss compared to a topology-specific training.

\subsection{Influence of global graph features}
\label{app:global_feature_ablation}

\begin{figure}[htbp]
    \centering
    \includegraphics[width=\columnwidth]{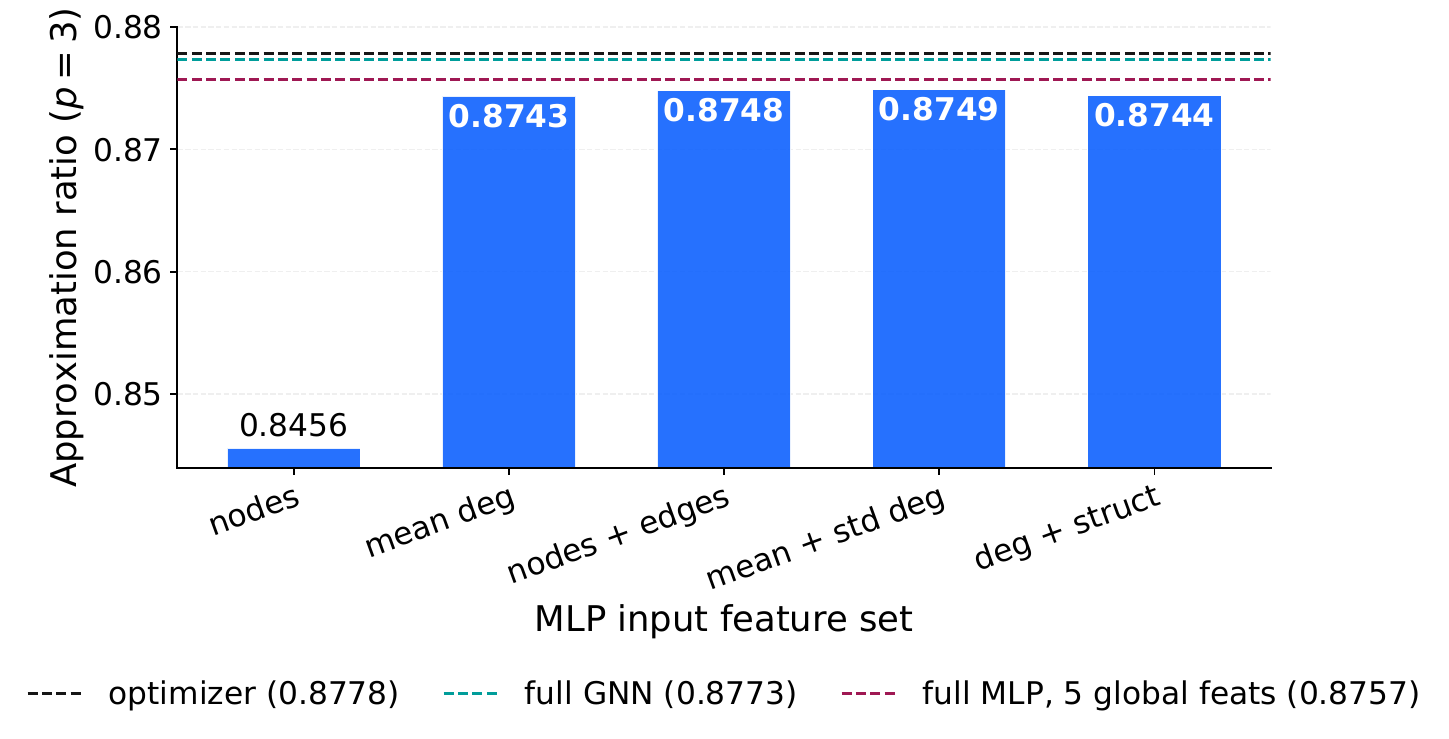}
    \caption{
    MLP input-feature ablation at $p=3$ on the multi-topology test set.
    Bars show the mean approximation ratio for models trained with different subsets of global graph features.
    Dashed lines indicate the approximation ratios of the solver, the full GNN, and the full MLP.
    }
    \label{fig:mlp_feature_ablation}
\end{figure}

The MLP predicts QAOA angles only from global graph statistics and nevertheless achieves a strong mean performance.
To understand this result, we ablate the input features at $p=3$.
Using only the number of nodes reduces the AR, see Fig.~\ref{fig:mlp_feature_ablation}.
However, the mean degree alone recovers most of the performance of the full MLP.
Adding graph size or degree variation provides a small further improvement. Additional structural statistics do not improve over the simpler degree-based feature sets.
This indicates that degree information explains a large part of the predictive signal available to the global-feature~MLP.

\section{Distribution-shift experiments}
\label{app:distribution_shift}

As discussed in the main text, we evaluate generalization beyond the training distribution in both graph size and topology. Here, we provide additional details on these graph instances and the classically optimized QAOA angles used as reference solutions. In addition, we further discuss the implications of our findings in the out-of-distribution experiments.

\subsection{Large-scale graphs}

\begin{figure}[tbp]
    \centering
    \includegraphics[width=1\linewidth]{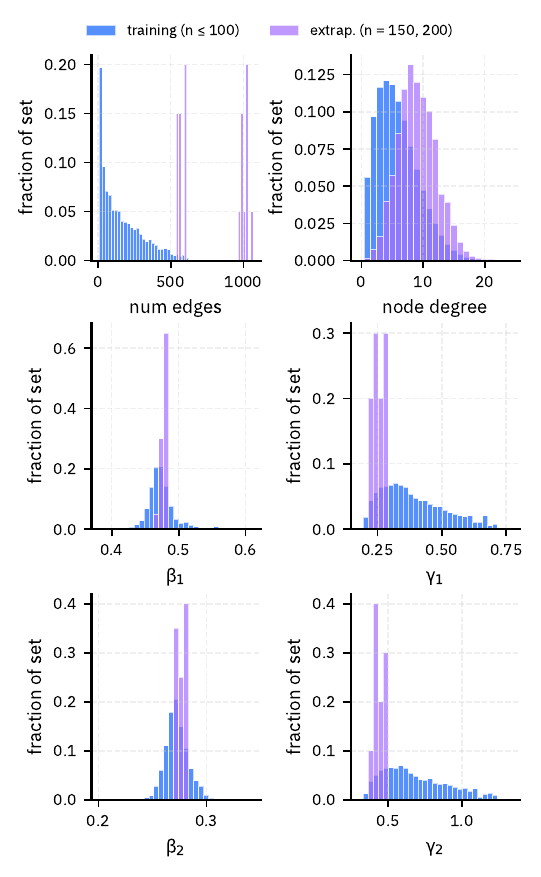}
    \caption{Distribution of graph properties and optimized QAOA angles at depth $p=2$ for the training set ($n\leq100$, blue) and the larger-graph extrapolation set ($n\in\{150,200\}$, purple). Shown are the number of edges, node-degree distribution, and the optimized angles $(\beta_1,\gamma_1,\beta_2,\gamma_2)$. Histograms are normalized separately to show the fraction of instances in each set.}
    \label{fig:large-graphs-distribution}
\end{figure}

To assess extrapolation to unseen graph sizes, we evaluate the learned models on graphs with 150 and 200 nodes. 
These instances lie beyond the graph sizes in the training set.  
The distributions of the $\boldsymbol{\gamma}$ angles of the in-distribution Erd\H{o}s--R\'enyi graphs used for training in the main paper differ significantly from the distributions of the $\boldsymbol{\gamma}$ angles of the out-of-distribution Erd\H{o}s--R\'enyi graphs used in this experiment, see Fig.~\ref{fig:large-graphs-distribution}. 
In addition, we notice different graph characteristics between the in-distribution Erd\H{o}s--R\'enyi graphs and the larger-scale, out-of-distribution Erd\H{o}s--R\'enyi graphs as the graph-level statistics differ. 
This causes a distribution shift in the input to the deep learning models.
Therefore, we expect that the inferred angles of the deep learning models may have a sub-optimal performance.
We confirm this hypothesis in the main text, where we report a decline in the AR.
This finding leaves room for two research directions. 
(i)~Continuously train models on larger graph instances at the cost of generating more training data. 
(ii)~Decompose larger graphs into sub-graphs which fit the current models and re-compose for a final set of angles. 
We expect that this re-composition processes could also benefit from deep learning.  

\subsection{Unseen graph topologies\label{app:unseen_topology}}

\begin{figure}[tbp]
   \centering
   \includegraphics[width=1\linewidth]{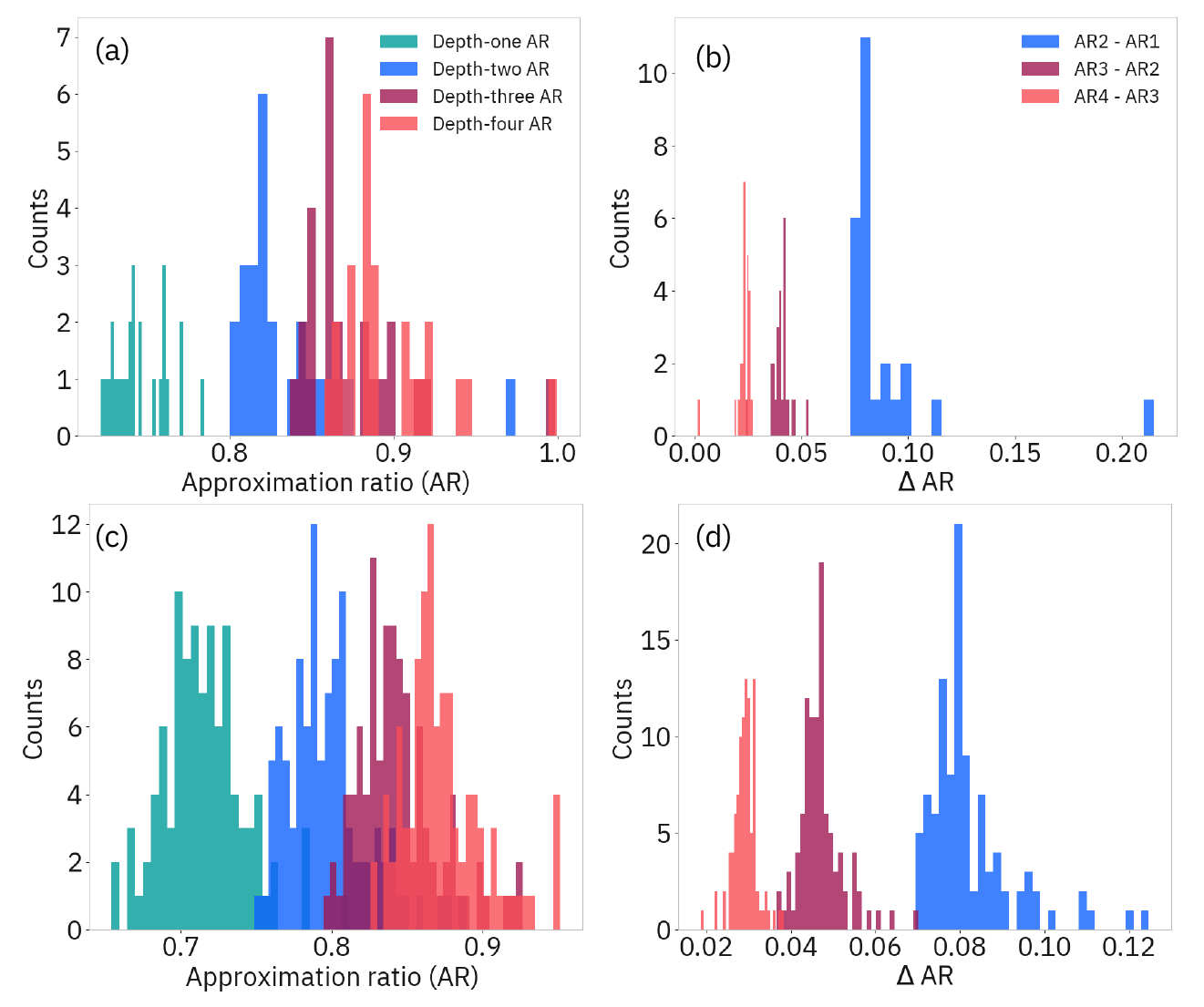}
    \caption{\textsc{MaxCut} performance and depth-wise scaling on weighted Barabási–Albert and 2D grid graphs. Panels (a) and (c) are histograms of approximation ratios $\alpha_p$ for circuit depths $p=1$ (teal), $p=2$ (blue), $p=3$ (magenta), and $p=4$ (coral). Panels (b) and (d) are empirical distributions of consecutive layer gains $\Delta \alpha =\hat{\alpha}_{p} - \hat{\alpha}_{p-1}$.}
    \label{fig:ar_dist_ba_2d}
\end{figure}

In the main text, we evaluate how our models generalize to weighted
Barab\'asi--Albert and 2D grid graphs.
Neither appear in the training set.
The distributions of the \emph{solver} optimized QAOA angles $\boldsymbol{\beta}$ and $\boldsymbol{\gamma}$ for Barab\'asi--Albert and 2D grid instances are concentrated within $[0,\pi/2]$ and are unimodal, with no secondary clusters or $\pi$-shifts (data not shown). 
The angles follow a smooth schedule across the QAOA layers. 
This resembles the monotone parameter schedules reported for QAOA and their connection to adiabatic evolution and is a consequence of the Interp. training~\cite{Zhou2020}.
The approximation ratio distributions shift to higher values as $p$ increases from $1$ to $4$, see Fig.~\ref{fig:ar_dist_ba_2d}(a) and~(c). 
The gains in AR, i.e., $\Delta\textrm{AR} = \textrm{AR}(p) - \textrm{AR}(p-1)$, are non-negative for all evaluated instances, see Fig.~\ref{fig:ar_dist_ba_2d}(b) and~(d).
Therefore, increasing $p$ never reduces the AR, as expected. 
Overall, the strong unimodal concentration of angles in $[0, 2\pi]$ benefits models during extrapolation as even when input distributions change, the targets are aligned with the distribution of the training targets.

\begin{table*}
\centering
\caption{Performance metrics (approximation ratio and reach, \%, median (IQR)) averaged over 5 independent runs across 4 quantum circuit depths ($p = 1 \ldots 4$) and 7 model architectures.}
\vspace{0.6em}
\label{tab:depth_architecture_summary}
\begin{tabular}{clcccc}
\hline\hline
 &  & \multicolumn{2}{c}{\textbf{Barabasi-Albert}} & \multicolumn{2}{c}{\textbf{2D Grids}} \\
\cmidrule(lr){3-4} \cmidrule(lr){5-6}
\textbf{Depth} & \textbf{Architecture} & \textbf{AR} & \textbf{Reach} & \textbf{AR} & \textbf{Reach} \\
\midrule
1 & GNN & 70.96 (3.74) & 99.56 (1.30) & 74.54 (1.00) & 99.97 (0.03) \\
 & GIN & 71.21 (3.07) & 99.63 (0.42) & 74.45 (1.02) & 99.85 (0.10) \\
 & GCN & 71.23 (2.94) & 99.68 (0.55) & 74.50 (1.02) & 99.91 (0.06) \\
 & DiffTF & 71.08 (3.06) & 99.42 (0.56) & 73.88 (1.17) & 99.05 (0.52) \\
 & EdgeTF & 71.24 (2.83) & 99.62 (0.36) & 74.50 (1.01) & 99.93 (0.05) \\
 & GraphTF & 71.26 (2.80) & 99.61 (0.45) & 74.38 (1.02) & 99.77 (0.08) \\
 & MLP & 71.37 (2.81) & 99.78 (0.27) & 74.53 (0.99) & 99.96 (0.03) \\
 & Fixed Angle & 66.17 (3.43) & 92.50 (2.61) & 66.96 (0.51) & 90.01 (0.97) \\
 & Solver & 71.59 (-) & - & 74.57 (-) & - \\
\midrule
2 & GNN & 67.12 (24.05) & 83.96 (29.37) & 83.36 (1.35) & 99.61 (0.13) \\
 & GIN & 78.24 (3.31) & 98.51 (2.01) & 83.11 (1.35) & 99.38 (0.28) \\
 & GCN & 79.37 (2.41) & 99.57 (0.72) & 82.80 (1.31) & 99.00 (0.53) \\
 & DiffTF & 77.31 (3.14) & 96.92 (2.07) & 80.74 (1.66) & 96.41 (0.94) \\
 & EdgeTF & 79.05 (2.57) & 99.16 (0.97) & 83.18 (1.31) & 99.40 (0.23) \\
 & GraphTF & 78.75 (2.79) & 98.77 (1.50) & 82.85 (1.35) & 99.12 (0.26) \\
 & MLP & 79.25 (2.56) & 99.45 (0.78) & 83.24 (1.30) & 99.49 (0.11) \\
 & Fixed Angle & 69.17 (3.36) & 86.32 (3.00) & 71.24 (0.73) & 85.23 (1.03) \\
 & Solver & 79.86 (-) & - & 83.72 (-) & - \\
\midrule
3 & GNN & 69.07 (24.44) & 81.83 (28.95) & 87.07 (1.63) & 99.48 (0.71) \\
 & GIN & 80.63 (4.42) & 95.65 (3.61) & 86.89 (1.31) & 99.03 (0.45) \\
 & GCN & 81.43 (3.59) & 96.43 (3.07) & 84.54 (2.17) & 96.35 (2.12) \\
 & DiffTF & 81.35 (3.14) & 96.05 (2.17) & 83.59 (2.03) & 95.33 (1.52) \\
 & EdgeTF & 82.83 (2.52) & 97.81 (1.50) & 85.71 (1.65) & 97.73 (0.93) \\
 & GraphTF & 82.76 (2.63) & 97.76 (1.70) & 86.01 (1.74) & 98.24 (1.03) \\
 & MLP & 83.51 (2.57) & 98.56 (1.29) & 84.41 (1.94) & 96.12 (1.39) \\
 & Fixed Angle & 69.01 (4.11) & 82.01 (4.21) & 72.83 (0.98) & 82.88 (1.03) \\
 & Solver & 84.49 (-) & - & 87.75 (-) & - \\
\midrule
4 & GNN & 67.81 (31.30) & 77.72 (36.32) & 88.90 (1.69) & 98.98 (0.93) \\
 & GIN & 82.88 (4.32) & 94.97 (4.19) & 88.23 (1.42) & 98.23 (0.60) \\
 & GCN & 84.77 (3.33) & 96.96 (2.40) & 87.07 (2.01) & 97.18 (1.50) \\
 & DiffTF & 82.59 (3.71) & 94.28 (3.07) & 85.46 (2.08) & 94.94 (1.52) \\
 & EdgeTF & 85.52 (2.81) & 97.44 (1.70) & 86.91 (1.82) & 96.72 (1.21) \\
 & GraphTF & 85.52 (2.74) & 97.77 (1.90) & 88.10 (1.58) & 97.99 (0.67) \\
 & MLP & 86.20 (2.73) & 98.37 (1.35) & 84.82 (1.90) & 94.35 (1.38) \\
 & Fixed Angle & 67.98 (4.00) & 78.03 (3.87) & 71.85 (1.01) & 79.75 (1.16) \\
 & Solver & 87.36 (-) & - & 89.99 (-) & - \\
\hline\hline
\end{tabular}
\end{table*}

\section{Model architecture details}
\label{app:model_details}

This section provides additional implementation details for the seven model architectures introduced in Sec.~\ref{sec:ml_models}. Following the organization in the main text, we consider a non-structural MLP baseline and six graph-based models grouped into three classes: edge-token transformers (EdgeTF and DiffTF), node-based graph encoders (GraphTF and GCN), and message-passing graph networks (GNN and GIN). The models differ primarily in how they represent and process the graph structure and edge weights before mapping the resulting graph representation to the QAOA angles. 
Let $G(V, E)$ be a graph with nodes $v_i\in V$ and edges $(v_i,v_j)\in E$, and let $p$ denote the QAOA circuit depth. For each depth, a separate model is trained to predict the variational parameters $\theta = (\beta_1,\ldots,\beta_p, \gamma_1,\ldots,\gamma_p) \in \mathbb{R}^{2p}$ that minimize the QAOA expected energy on the Max-Cut cost Hamiltonian.

\subsection{Multilayer Perceptron}
As a non-structural baseline, we use a Multilayer Perceptron (MLP), a standard feedforward neural network composed of fully connected layers~\cite{goodfellow2016deep}, that operates exclusively on the global graph-level features. The model consists of four hidden layers, each comprising a linear transformation followed by a $\text{ReLU}$ activation and dropout regularization. Layer normalization is applied after the final hidden layer, followed by a linear projection to the $2p$ QAOA angles. Since the MLP does not use the graph structure or edge weights, it provides a reference for how well the angles can be predicted from coarse global graph properties alone.

Let $h^{(0)} \in \mathbb{R}^{d_{\text{in}}}$ denote the input feature vector. For each hidden layer $l \in \{1, \dots, L\}$, the intermediate representation $h^{(l)}$ is computed as:

\begin{equation}
    h^{(l)} = \mathrm{Dropout} \left(\sigma\left(W^{(l)} h^{(l-1)} + b^{(l)}\right)\right),  
\end{equation}

where $W^{(l)} \in \mathbb{R}^{d_l \times d_{l-1}}$ and $b^{(l)} \in \mathbb{R}^{d_l}$ denote the weight matrix and bias vector, respectively, and $\sigma(\cdot)$ denotes the $\text{ReLU}$ activation function. Layer normalization is applied after the final hidden layer, followed by a linear output projection,
\begin{equation}
\hat{\boldsymbol{\theta}}
=
W_{\mathrm{out}}
\mathrm{LayerNorm}\left(h^{(L)}\right)
+
b_{\mathrm{out}},
\end{equation}
where $\hat{\boldsymbol{\theta}} \in \mathbb{R}^{2p}$ denotes the predicted QAOA angles.

\subsection{Edge-token transformer models}
Edge-token transformer models represent each graph edge as a separate token and process all edge tokens jointly using self-attention. Each edge token encodes information about its two endpoint nodes and the corresponding edge weight. A global token constructed from graph-level features is processed together with the edge tokens and provides the graph-level representation used to predict the QAOA angles.
We consider two models in this category: the Edge Transformer (EdgeTF) and the diffusion-based Edge Transformer (DiffTF). DiffTF extends EdgeTF by introducing diffusion-inspired noise conditioning during training.

\subsubsection{Edge transformer}

The edge-token transformer (EdgeTF) uses a standard Transformer encoder~\cite{vaswani2017attention} to process the edge and global tokens. This enables interactions between all graph edges through self-attention, rather than restricting information exchange to local neighborhood updates. In addition to the edge tokens, a global token is constructed from the global graph features and processed jointly with the edges.

The graph-level feature vector $\mathbf{f} \in \mathbb{R}^F$, containing the number of nodes $|V|$, number of edges 
$|E|$, mean degree $\bar{d}$, and degree standard deviation $\sigma_d$, is projected into the Transformer embedding space using a single 
linear layer to produce a global token $\mathbf{t}_{\text{cls}} \in \mathbb{R}^d$.

For each edge $(u, v) \in E$, the two endpoint node indices are independently encoded using sinusoidal positional encodings~\cite{vaswani2017attention}, producing 
vectors $\mathbf{e}_u, \mathbf{e}_v \in \mathbb{R}^{d_e}$. To obtain an endpoint-order-invariant edge representation,, the mean and absolute difference of the two endpoint encodings are concatenated with the scalar edge weight $w_{uv}$,
\begin{equation}
    \mathbf{r}_{uv} = \left[\,\frac{\mathbf{e}_u + \mathbf{e}_v}{2}\ \Bigg\|\ 
    |\mathbf{e}_u - \mathbf{e}_v|\ \Bigg\|\ w_{uv}\,\right] \in \mathbb{R}^{2d_e+1}
\end{equation}
which is then linearly projected to $\mathbf{t}_{uv} \in \mathbb{R}^d$, yielding one edge token per edge. The global token and all $|E|$ 
edge tokens are concatenated into a sequence of length $1 + |E|$ and passed through an $L$-layer Transformer encoder with $H$ attention 
heads and feedforward width $2d$, where multi-head self-attention allows every token to attend to every other token simultaneously, 
capturing pairwise and higher-order structural dependencies across the full edge set. After the final Transformer layer, the updated global token $\mathbf{z} \in \mathbb{R}^d$, enriched through attention over the edge tokens, is used as the graph-level representation and decoded by a final linear layer to predict the QAOA angles,
\begin{equation}
    (\boldsymbol{\gamma}, \boldsymbol{\beta}) = \mathbf{W}_o\,\mathbf{z} + \mathbf{b}_o 
    \in \mathbb{R}^{2p}
\end{equation}
Since the endpoint encodings depend on the node indices, we randomly relabel the nodes during training before constructing the edge tokens. This prevents the model from learning fixed node identities and encourages predictions that are insensitive to the arbitrary labeling of the graph. We use sinusoidal rather than learned node-index embeddings, which also allows the same encoding to be applied to graphs with more nodes than those encountered during training.

Our edge-token representation is conceptually related to the Edge Transformer of~\citet{bergen2021edge}, which represents relations between node pairs directly rather than relying only on node representations. In contrast to their architecture, which maintains representations for all node pairs and uses triangular attention, our EdgeTF represents only the observed graph edges and processes them using standard self-attention.

\subsubsection{Diffusion-based edge-token transformer}

The diffusion-based edge-token transformer (DiffTF) extends EdgeTF by introducing diffusion-inspired noise conditioning during training, based on the forward diffusion process of Denoising Diffusion Probabilistic Models (DDPMs)~\cite{ho2020denoising}. 
In DDPM, data are gradually corrupted with Gaussian noise according to a predefined noise schedule and the model is trained to learn the reverse denoising process through variational inference. In our implementation, we use the forward noising process as a stochastic regularizer, while retaining the deterministic regression setting of EdgeTF.

We apply a linear noise schedule $\{\beta_t\}_{t=1}^{T}$ that interpolates from $\beta_{\min}$ to $\beta_{\max}$ over $T$ timesteps. The cumulative signal-retention coefficients are defined as $\bar{\alpha}_t = \prod_{s=1}^{t}(1-\beta_s)$, and the forward diffusion process corrupts the input according to

\begin{equation}
    \mathbf{x}_t = \sqrt{\bar{\alpha}_t}\,\mathbf{x}_0
                 + \sqrt{1 - \bar{\alpha}_t}\,\boldsymbol{\epsilon},
    \quad \boldsymbol{\epsilon} \sim \mathcal{N}(\mathbf{0},\mathbf{I}),
    \label{eq:forward_diffusion}
\end{equation}

where, during training, a timestep $t \sim \mathrm{Uniform}\{1,\ldots,T\}$ is sampled and the resulting noised input 
$\mathbf{x}_t$ is used. At inference, the model acts as a 
deterministic regressor on the clean input ($t=0$). Thus, in contrast to a generative DDPM, no iterative reverse-diffusion chain is performed.

Given a graph $G=(V,E,\mathbf{w})$, the noised features are first projected into a latent space of dimension $d$ via 
a linear transformation, yielding $\mathbf{h}=\mathbf{W}_{\mathrm{in}}\mathbf{x}_t$. 
A sinusoidal timestep embedding  $\phi(t)\in\mathbb{R}^{d}$~\citep{vaswani2017attention} is processed by a two-layer SiLU MLP and added to the projected features to provide information about the noise level. The 
conditioned representation is then passed to the same edge-token Transformer backbone as EdgeTF.

As described above for EdgeTF, an edge token is constructed for every edge $(u,v)\in E$ from sinusoidal encodings of the endpoint indices, their mean and absolute difference, and the corresponding edge weight $w_{uv}$. A global token represents the conditioned graph-level features. The global token and edge tokens are then concatenated into a sequence 
and processed by an $L$-layer, $H$-head Transformer encoder. The output corresponding to the global 
token is taken as the graph-level representation. Finally, the graph 
representation is passed through layer normalization, dropout, and a linear output layer to predict 
the QAOA angles $\hat{\boldsymbol{\theta}} \in \mathbb{R}^{d_{\mathrm{out}}}$. 

As for EdgeTF, node indices are randomly relabeled during training before constructing the edge tokens. Sinusoidal rather than learned node-index embeddings are used, allowing the same encoding to be evaluated for node indices beyond those encountered during training. 

\subsection{Node-based graph encoders}

Unlike the edge-token models, node-based graph encoders represent the graph through node features that are updated using the graph connectivity and edge weights. The resulting node representations are pooled into a graph-level representation and combined with global graph features to predict the QAOA angles. We consider two models in this category: the Graph Transformer (GraphTF) and the Graph Convolutional Network (GCN).

\subsubsection{Graph Transformer}

The Graph Transformer (GraphTF) processes the node representations using multi-head self-attention, with attention restricted to neighboring nodes according to the graph connectivity. Edge weights are incorporated through an additional learned bias in the attention scores.

In our implementation, the nodes are initialised with two scalar features: the log-transformed node degree and signed-log-transformed weighted-degree. These are linearly projected to higher dimensional space to obtain the initial node embeddings,

\begin{equation}
    h_i^\text{(0)} = W_\text{node} \cdot f_i 
\end{equation}
\

where $W_{\text{node}}$ is the learned projection matrix and $f_i$ is the two-dimensional feature vector of node $i$. Laplacian positional encodings are computed using the symmetric normalised Laplacian~\cite{dwivedi2021generalizationtransformernetworksgraphs},

\begin{equation}
    \text{L}=\text{I} - \text{D}^{ - 1/2}\text{A}\text{D}^{ - 1/2}
\end{equation}
where $A$ is the binary adjacency matrix and $D$ the corresponding degree matrix. The eigenvectors corresponding to the smallest nontrivial eigenvalues are used as positional encodings. To address the inherent sign ambiguity of the Laplacian eigenvectors, each eigenvector is multiplied by a random sign (\verb|+1| or \verb|-1|) with equal probability \cite{lim2022signbasisinvariantnetworks,kreuzer2021rethinkinggraphtransformersspectral} during training. The positional encodings are then linearly projected and added to the initial node embeddings,

\begin{equation}
    h_i^{(0)} \leftarrow h_i^{(0)} + W_{pos} \cdot p_i
\end{equation}

The resulting node embeddings are passed through Transformer layers with edge-masked multi-head self-attention. For attention head $h$, the scaled dot-product attention score is 
\begin{equation}
\textit{e}_{ij} = \frac{(W_Qh_i)(W_Kh_j)^T}{\sqrt{d_{head}}}
\end{equation}
Attention scores for non-neighboring nodes are set to ${-\infty}$. This effectively reduces to an attention weight of zero, which ultimately restricts attention to only node neighbors \cite{dwivedi2021generalizationtransformernetworksgraphs}. If edge weights are available, a per-attention-head additive bias is produced and added to the attention scores:

\begin{equation}
    \textit{e}_{ij}^{(h)} \leftarrow \textit{e}_{ij}^{(h)} + MLP(signedlog1p(\textit{w}_{ij}))^{(h)}
\end{equation}
where $e_{ij}^{(h)}$ is the attention score between node \textit{i} and node \textit{j} for attention head \textit{h} specifically. $signedlog1p(w_{ij})$ is compression of edge weight \textit{ij} which is passed on to a small learned MLP to produce one scalar bias per attention head $MLP(\cdot)^{(h)}$ \cite{ying2021transformersreallyperformbad}. After the final layer, the node representations are mean-pooled into a graph-level vector, concatenated with the projected global graph features, and passed through an MLP to predict the QAOA angles.

\subsubsection{Graph Convolutional Network}    
As a standard node-based graph architecture, we include a graph convolutional network based on the \texttt{GraphConv} operator~\cite{morris2019weisfeiler}. The model represents each node by a feature vector and iteratively updates this representation using information from directly connected nodes. For our weighted graphs, the contribution from each neighboring node is scaled by the corresponding edge weight. 
Formally, for a graph with $N$ nodes, feature matrix $X \in \mathbb{R}^{N \times F}$, and edge weights $w_{ji}$, the layer-wise update for node $i$ is given by

\begin{equation}
    h_i^{(l)}
    =
    W_{\mathrm{self}}^{(l)} h_i^{(l-1)}
    +
    W_{\mathrm{neigh}}^{(l)}
    \sum_{j \in \mathcal{N}(i)}
    w_{ji} h_j^{(l-1)},
    \label{eq:graphconv_message_pass}
\end{equation}

where $h_i^{(l)}$ denotes the representation of node $i$ at layer $l$, with $H^{(0)}=X$, and $W_{\mathrm{self}}^{(l)}$ and $W_{\mathrm{neigh}}^{(l)}$ are trainable layer-specific weight matrices. Each layer therefore combines the current representation of a node with a weighted sum of the representations of its immediate neighbors. We group this model with the node-based graph encoders, since each layer updates the node representations through a direct weighted aggregation over neighboring nodes. The message-passing graph networks introduced below differ in that the information passed along each edge is itself learned before aggregation.

For our implementation, the initial node features are given by the one-dimensional node degree vector, $z$-score normalized within each graph. 
The network consists of two such \texttt{GraphConv} layers, each followed by a ReLU activation and dropout regularisation. A global mean-pooling readout then aggregates the node embeddings into a single, fixed-size graph representation. This graph representation is concatenated with global graph-level features before being passed to a two-layer prediction head that outputs the $2p$ QAOA angles.

\subsection{Message-passing graph networks}

Message-passing graph networks update node representations by exchanging and aggregating information along graph edges~\cite{corso2024graph,gilmer2017neural}. In our implementations, this involves three steps: message computation, neighborhood aggregation, and node-state updates. We consider two models in this class: a Graph Neural Network (GNN) and a Graph Isomorphism Network (GIN).

\subsubsection{Graph Neural Network}
Graph Neural Networks (GNNs) extend deep learning to graph-structured data by iteratively refining node representations. By combining structural connectivity with feature attributes, they learn representations at the node, edge, and graph levels. Under the Message-Passing Neural Network (MPNN) framework~\cite{gilmer2017neural}, a GNN updates node states across $L$ layers through local neighborhood aggregation. At layer $l \in \{1, \dots, L\}$, the update for node $v_i$ is governed by three functions:
\begin{align}
    \mathbf{m}_{ij}^{(l)} =&\, \text{MESSAGE}^{(l)} \left( h_i^{(l-1)}, h_j^{(l-1)}, e_{ij} \right)
    \label{eq:gnn_message_step}\\
    \mathbf{a}_i^{(l)} =&\, \text{AGGREGATE}_{j \in \mathcal{N}(i)}^{(l)} \left( \mathbf{m}_{ij}^{(l)} \right)
    \label{eq:gnn_aggregate_step}\\
    h_i^{(l)} =&\, \text{UPDATE}^{(l)} \left( h_i^{(l-1)}, \mathbf{a}_i^{(l)} \right)
    \label{eq:gnn_update_step}
\end{align}
where $\mathcal{N}(i)$ is the set of 1-hop neighbors of node $v_i$. Here $i$ is the receiving node and $j$ ranges over its neighbors.

The output layer depends on the task. A node-level task projects the final node embeddings directly. An edge-level task combines pairs of node embeddings. A graph-level task applies a permutation-invariant pooling operation over all nodes $V$.

Let $N$ be the number of nodes and $X \in \mathbb{R}^{N \times F}$ the initial node feature matrix. We construct $X$ by projecting global graph-level statistics into the node embedding space. Every node receives the same context vector.

In each message-passing layer, we scale the neighbor's embedding element-wise by the edge weight, concatenate it with the receiver's embedding, and pass the result through the message MLP. Equation~(\ref{eq:gnn_message_step}) thus becomes
\begin{equation}
    \mathbf{m}_{ij}^{(l)} = \text{MLP}_{message} \left( h_i^{(l-1)} \parallel \tilde{h}_j^{(l-1)} \right)
\end{equation}
where $\tilde{h}_j^{(l-1)} = w_{ij} \cdot h_j^{(l-1)}$ is the edge-weighted neighbor embedding.

Conditioning the message on the receiver embedding $h_i$ is a deliberate design choice. The message depends on what the receiver already knows, which gives a form of attention-like selectivity without an explicit attention mechanism. Edges arriving at different nodes can therefore produce different messages. The message MLP is a two-layer network with a ReLU in between.

Nodes then sum the incoming messages, so Equation~(\ref{eq:gnn_aggregate_step}) becomes
\begin{equation}
    \mathbf{a}_i^{(l)} = \sum_{j \in \mathcal{N}(i)} \mathbf{m}_{ij}^{(l)}.
\end{equation}
This is a plain sum with no degree normalization. High-degree nodes accumulate larger aggregates. This makes the representation sensitive to local connectivity density, which is relevant for QAOA because high-degree nodes appear in more terms of the cost Hamiltonian.

Finally, each node is updated by concatenating its current embedding with the aggregated neighborhood signal and passing both through an MLP. Equation~(\ref{eq:gnn_update_step}) becomes
\begin{equation}
    h_i^{(l)} = \text{MLP}_{update} \left( h_i^{(l-1)} \parallel \mathbf{a}_i^{(l)} \right).
\end{equation}
Like the message MLP, the update MLP is a two-layer network with a ReLU after the first layer.

Among our message-passing models, the GNN is the only one whose message function uses both endpoints of an edge. The network stacks $L$ such message-passing layers. A global mean-pooling readout then aggregates the node embeddings into a fixed-size graph representation. This representation is concatenated with the raw graph-level features and passed to a prediction head with dropout, which outputs the QAOA parameters.

\subsubsection{Graph Isomorphism Network}

The GIN was proposed by Xu \emph{et al.}~\cite{xu2018powerful} to address the non-injective aggregation and pooling of standard GCNs. The authors show that any message-passing GNN is at most as expressive as the 1-dimensional Weisfeiler-Leman (1-WL) graph isomorphism test. They design the GIN to reach this bound by making both its update and its readout injective. Unlike GCN or GraphSAGE, which aggregate neighbors with mean or max pooling, the GIN uses sum aggregation, which the authors prove is a universal, injective function over multisets. At layer $\ell$, node $u$ is updated as
\begin{equation}
    h_u^{(\ell)} = \phi \Big( \big( 1 + \epsilon \big) \, h_u^{(\ell-1)}
    + \sum_{v \in \mathcal{N}(u)} h_v^{(\ell-1)} \Big)
\end{equation}
where $\phi$ is a parameterized MLP, $\epsilon$ is a fixed or learnable self-loop scaling parameter, and the sum aggregates the neighborhood. For an injective graph-level readout, the authors further propose concatenating the summed node representations from every layer,
\begin{equation}
    h_G = \text{CONCAT} \left( \sum_{v \in V} h_v^{(0)},\
    \sum_{v \in V} h_v^{(1)},\ \dots,\ \sum_{v \in V} h_v^{(L)} \right),
\end{equation}
so that structurally distinct graphs cannot map to identical embeddings.

The equations above state the conditions that make a GIN as powerful as the 1-WL test. Our implementation follows this design but adapts it in two practical ways, described below.

First, we condition messages on edge weights. Before summation, each neighbor embedding is scaled by its scalar edge weight, so the aggregation becomes $\sum_{v \in \mathcal{N}(u)} w_{uv} \, h_v^{(\ell-1)}$. We refer to these as edge-conditioned GIN layers.

Second, we use a single global pooling step rather than the layer-wise concatenated readout above. By default we pool node embeddings with a mean operation. The sum-based, injective readout is therefore available but not the default configuration in our experiments.

The GIN regressor takes as input a graph $G = (V, E)$ representing a QAOA instance, together with a graph-level feature vector $\mathbf{f} \in \mathbb{R}^F$ of global statistics. These statistics are the number of nodes $|V|$, the number of edges $|E|$, the edge-to-node ratio $|E|/|V|$, the mean degree $\bar{d}$, and the degree standard deviation $\sigma_d$. 
Since nodes carry no explicit attributes, $\mathbf{f}$ is first projected through a two-layer ReLU MLP into a global embedding $\mathbf{g} \in \mathbb{R}^d$. This embedding is broadcast to all $|V|$ nodes and optionally concatenated with pre-computed per-node structural features, such as the degree z-score. A second two-layer ReLU MLP then fuses these into a uniform initial node embedding $\mathbf{h}_v^{(0)} \in \mathbb{R}^d$ for every $v \in V$.

The embeddings are refined through $L$ stacked edge-conditioned GIN layers. Each layer applies a residual connection followed by Layer Normalization, which together prevent representational collapse at depth. After the final layer, a pre-pooling projection MLP is applied and the node embeddings are pooled into a graph-level vector $\mathbf{z} \in \mathbb{R}^d$. Finally, $\mathbf{z}$ is concatenated with $\mathbf{f}$ to form $[\mathbf{z} \| \mathbf{f}] \in \mathbb{R}^{d+F}$, which a three-layer output MLP with dropout decodes into the predicted QAOA parameters $(\boldsymbol{\gamma}, \boldsymbol{\beta}) \in \mathbb{R}^{2p}$.

\end{document}